\documentclass[prb,twocolumn,showpacs,preprintnumbers,amsmath,amssymb,floatfix]{revtex4-2}
\usepackage{graphicx}
\usepackage{dcolumn}
\usepackage{bm}
\usepackage{slashed}

\usepackage{trfsigns}
\usepackage{ulem}

\usepackage{tensor}
\usepackage{color}

\newcommand \be{\begin{eqnarray}}
\newcommand \ee{\end{eqnarray}}
\newcommand \ba{\begin{align}}

\newcommand {\V}[1]{{\bf #1}}

\newcommand {\p}[1]{\partial_{#1}}

\begin{document}
\title{Interplay of viscosity and surface tension for ripple formation by laser melting}
\author{K. Morawetz$^{1,2}$, S. Trinschek$^{1}$, E. L. Gurevich$^{1}$
}
\affiliation{$^1$M\"unster University of Applied Sciences,
Stegerwaldstrasse 39, 48565 Steinfurt, Germany}
\affiliation{$^2$International Institute of Physics- UFRN,
Campus Universit\'ario Lagoa nova,
59078-970 Natal, Brazil}
\begin{abstract}
A model for ripple formation on liquid surfaces exposed to an external laser or particle beam and a variable ground is developed. The external incident beam is hereby mechanically coupled to the liquid surface due to surface roughness. Starting from the Navier Stokes equation the coupled equations for the velocity potential and the surface height are derived in a shallow-water approximation with special attention to viscosity. The resulting equations obey conservation laws for volume and momentum where characteristic potentials for gravitation and surface tension are identified analogously to conservative forces. The approximate solutions are discussed in the context of ripple formation in laser materials processing involving melting of a surface by a laser beam. 
Linear stability analysis provides the formation of a damped wave modified by an interplay between the external beam, the viscosity, and the surface tension. The limit of small viscosity leads to damped gravitational and the limit of high viscosity to capillary waves. The resulting wavelengths are in the order of the ripples occurring in laser welding experiments hinting to the involvement of hydrodynamic processes in their origin. By discussing the response of the system to external periodic excitations with the help of Floquet multipliers, we show that the ripple formation could be triggered by a a periodically modulated external beam, e.g. appropriate repetition rates of an incident laser beam.
The weak nonlinear stability analysis provides ranges where hexagonal or stripe structures can appear. The orientation of stripe structures and ripples are shown to be dependent on the incident angle of the laser or particle beam where a minimal angle is reported. Numerical simulations confirm the findings and allow to describe the influence of variable grounds.
\end{abstract}

\pacs{
}
\maketitle

\section{Introduction}

Laser material processing can be accompanied by the formation of periodic structures at different length scales. These periodic stripes or ripples are usually treated as an unwanted effect increasing the surface roughness in laser-ablation processes and conditions have been worked out to avoid such instabilities \cite{TK94}. However sometimes such structures can be used to improve tribological properties of the surface \cite{BONSEtribological}, to colorize it \cite{Dusser}, to manipulate the laser light polarisation \cite{Beresna} or to increase the efficiency of water splitting \cite{Bialuschewski2018}.

\begin{figure}
 \includegraphics[width=3.2cm]{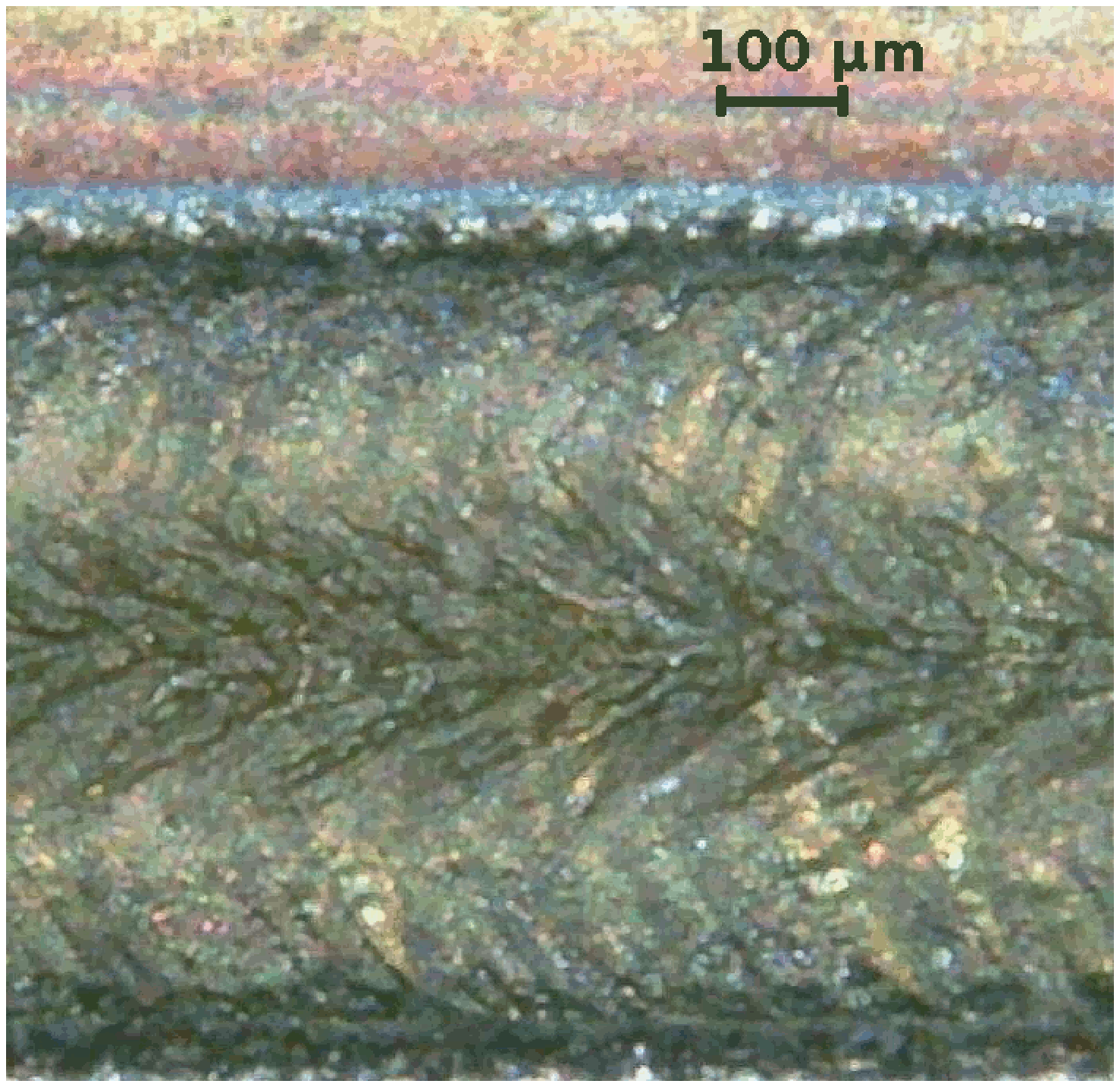} 
 \hspace{0.5cm}
\includegraphics[width=3.7cm]{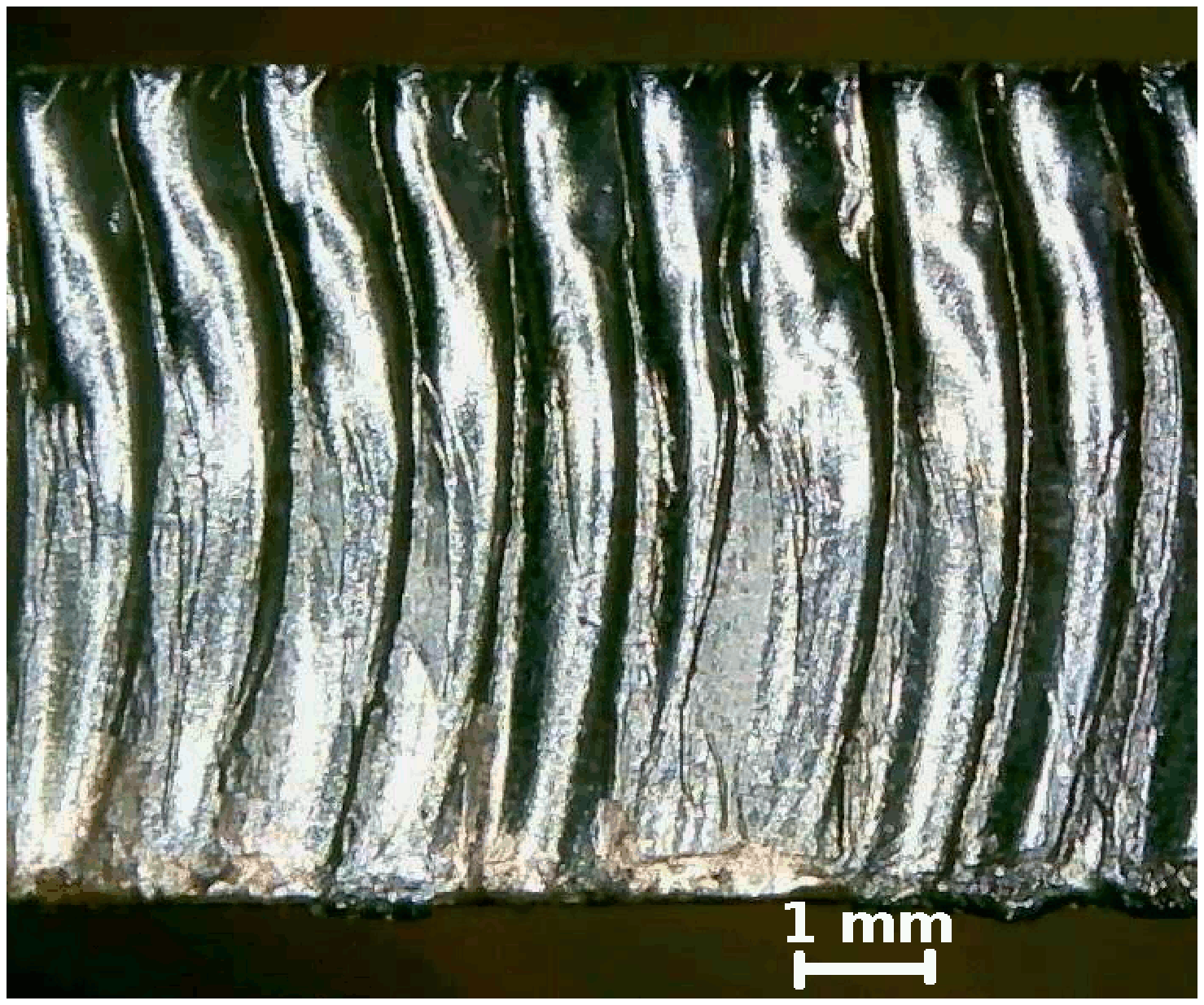}\\ 
\textbf{(a)}  \hspace{3.7cm}\textbf{(b)}\\
 \includegraphics[width=3.2cm]{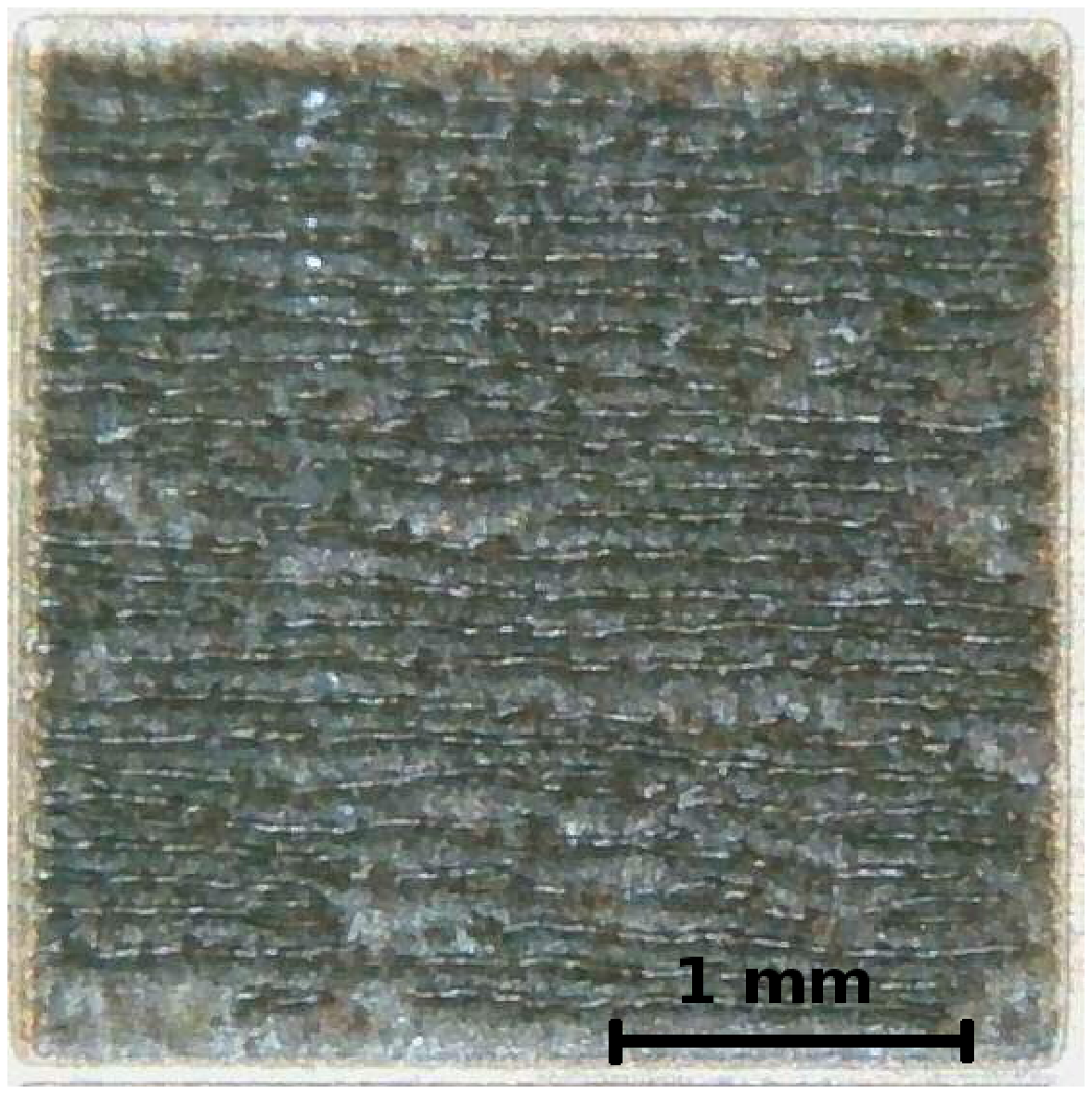} 
 \hspace{0.5cm}
 \includegraphics[width=3.8cm]{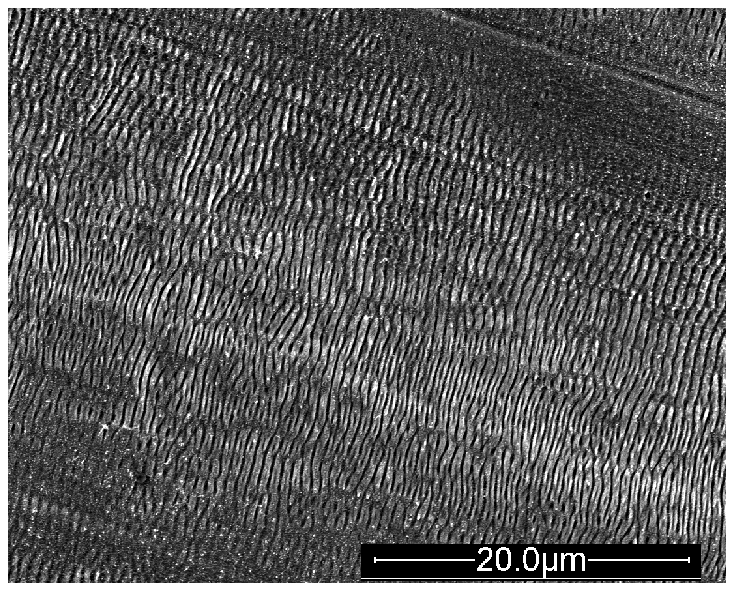} \\
\textbf{(c)}  \hspace{3.7cm}\textbf{(d)}
\caption{Laser-induced periodic structures: (a) weld sim on steel, welding with cw fiber laser, wavelength $\lambda=1.03\,\mu m$; (b) laser cut of a 6 mm thick steel with a cw diode laser $\lambda=0.8-1.0\,\mu m$, nitrogen atmosphere; (c) $3\times 3$mm area on steel engraved with a pulsed fiber laser, repetition rate $f=200\,kHz$, $\lambda=1.06\,\mu m$; (d) LIPSS left after femtosecond laser, repetition rate $f=1\,kHz$, $\lambda=0.8\,\mu m$.}\label{fig:exp}
\end{figure}

Periodic ripples seem to be generic for laser-solid interactions and can be observed in a large range of characteristic length scales on different surfaces and in different interaction regimes. Subwavelength ripples have been first seen by Birnbaum \cite{B65} on semiconductor surfaces processed with short-pulsed Q-switched ruby lasers. Ripple formation at larger length scales induced by hydrodynamic flow of molten metals and glasses induced by a long-pulsed CO$_2$ laser have been observed \cite{Siegrist1973,PDS87}. Periodic structures induced by a continuous-wave (cw) laser upon alloying were observed and analysed in \cite{AC77}. As examples, similar periodic ripples appear upon laser welding, see fig.~\ref{fig:exp}(a), laser cutting, see fig.~\ref{fig:exp}(b), engraving, see fig.~\ref{fig:exp}(c), or when a surface is exposed to femtosecond laser pulses and so-called LIPSS (Laser Induced Periodic Surface Structures) are formed, see fig.~\ref{fig:exp}(d). The common feature of all these cases is that the observed period of the structure is different from the laser wavelength and spot size and the laser light melts the surface.

Various mechanisms can lead to the phenomenon of periodic ripples. Periodic surface structures have been extensively investigated experimentally by femtosecond lasers to metals \cite{OPIKRYEI13,BonseRev,CLSFS14,MTDLNBOG17,FUENT19,FSFGSS19a} or semiconductors \cite{CE89,LPESS16} and by ion irradiation \cite{Chan:JAP2007,Chan:JAP2007,BVC17,NA19,Br20,LB20}. 
Three regimes of material response to femtosecond laser irradiation can be identified \cite{ZLI09,Ageev2016}: (i) melting and resolidification of a surface region of the target, (ii) photomechanical spallation
of a single or multiple layers or droplets, and (iii) an explosive disintegration of an overheated surface layer
as phase explosion. During the first regime of laser impact to a solid, surface melting occurs and surface instabilities can develop. This is accompanied by a periodic perturbation of the electronic temperature \cite{GLBN20} followed by an amplification, for given spatial periods, of the modulation in the lattice temperature \cite{GLGB17} and a final possible relocation by hydrodynamic instabilities. 

Models of ripples induced by cw or long pulse lasers are usually focused on hydrodynamic processes where the surface profile is created by the melt flow driven by temperature gradients at the surface \cite{AC77,LC81}. Surface oxidation is also sometimes discussed to be important in for chemically reactive metals like e.g., Ti processed at ambient atmosphere \cite{OPIKRYEI13}. Oxidation is observed on smooth and periodically patterned laser-processed surfaces, but as the ripples in fig.~\ref{fig:exp}(b) show they were obtained in $N_2$. 

To describe ripples induced by short laser pulses, the coupling of electromagnetism and hydrodynamics needs to be taken into account because the molten phase resolidifies fast. This reduces the time scale for hydrodynamic instabilities to grow. Early theoretical treatments for laser-induces ripples used electrodynamics to calculate the effective surface absorption of laser light on semiconductors to predict the occurring wavelength of ripples \cite{S83,GKKS89,CE89} which has been compared to experiments \cite{Y83}. In this context models for dielectric surfaces \cite{SKKV87} have considered the electric field produced by the induced polarization charge \cite{TS81}. The huge difference between the observed ripples wavelength in mm range and the laser wavelength requires a mechanism of down-converting. In \cite{EHW73} an interference between cavity and scattered radiation has been proposed and a multi-physical approach of combined Maxwell and hydrodynamical equations can be found in \cite{Rud19}. 

As mentioned by several authors before, the formation of periodic structures such as ripples or LIPSS are common not only for laser processing of solids, but also for electrical discharge erosion \cite{Henyk1999Dec} and ion beam etching \cite{COSTACHE2003}. There are also other abrasive surface processing methods like water jet cutting \cite{FriedrichPRL2000} which generates periodic ripples on the surface. This similarity encourages many authors to approach the modelling of the ripple formation starting not from the basic physical equations describing electrodynamic or hydrodynamic processes, like in \cite{AC77,LC81,S83}, but applying phemonenological models like Kardar-Parisi-Zhang and the Kuramoto-Sivashinsky equations \cite{FriedrichPRL2000,VarlamovaBestehorn2006,VRVB11}. In \cite{VRVB11}, the correlation of the LIPSS orientation with laser polarization has been investigated by assuming that the polarization causes a breaking of symmetry at the surface.

In this work, we focus on the regime of laser-induced ripples with hydrodynamical origin, i.e. ripples induced by cw or long pulses and laser welding. 
The liquid phase and the viscosity are assumed to be dominant compared to the transient optical properties as pronounced in \cite{GKEDM18}.  We will describe the regime of a melted surface and want to explore how far the phenomenon of ripple formation can be described by a mechanical origin. To that end, we present a purely hydrodynamical model that considers a laser or particle beam impact on a liquid bath as a (periodic) trigger of surface waves and nonlinear hydrodynamical processes followed by a freezing of the actual nonequilibrium situation.
At a larger time scale, thermal motion and Marangoni flows play a significant role \cite{BS15,KKRK16,GLGB17,SKKZRK13}. In this respect, thermo-capillary waves have been predicted \cite{LC81}. Thermal convection would be dominant if the Rayleigh number remains below $~10^3$ while values of $10^7$ are reported for key hole laser welding \cite{KDT11}.

We will treat here the regime of short-time processes before the main thermal convection will appear.	
Our model is based on a shallow water approximation of the Navier-Stokes equation in the presence of viscosity, surface tension and an external flow induced by a laser or particle beam.
In \cite{Dias08} a linearized theory of Navier-Stokes equation with damping by viscosity has been worked out but without surface tension. That the surface tension interferes crucially with the effect of viscosity has been demonstrated for capillary waves on the surface between ethanol and air \cite{AH87}. The dependence of the spatial period of ripples and the temperature has been calculated in \cite{GKEDM18}. Since the viscosity scales with the temperature one can turn the picture into a dependence of the ripple wavelength on the viscosity.

Here, we will show the interplay of viscosity and surface tension and will propose that the ripple formation can result from a hydrodynamical instability triggered by external incident beam with the help of Navier Stokes equation. 
 This will be reached by considering the coupling of the external beam to the surface roughness leading to a surface current. This is supported by the observation that surface roughness influences directly the period of structure \cite{FUENT19,FSFGSS19,FSFGSS19a}. From ion irradiation on surfaces, it is well established that surface roughness can induce pattern formation \cite{Chan:JAP2007}. The incident beam couples to the surface and results in ripples \cite{Shenoy:PRL2007} or nanodots \cite{Bradley:PRL2010}. During ion erosion, the pattern formation depends on the composition \cite{Shenoy:PRL2007,Bradley:PRL2010,Bradley:PRB2011,AM12}. Among these nanopatterning \cite{MG14} especially the ripple formation has been investigated \cite{BVC17}. For recent overviews see \cite{NA19,Br20,LB20}. Analogously we want to explore here the pattern formation on melted surfaces due to the laser-beam impact which facilitate hydrodynamic instabilities \cite{hydrodynamic2016}.  The dependence of ripple formation on the laser incident angle \cite{FUENT19} will be considered and it will be shown how the external laser beam creates ripples.

The outline of the paper is as follows. In the next chapter the liquid formulas are shortly reviewed together with the boundary conditions. Then the approximate equations are developed analogously to shallow-water equations but including external currents, surface tension and viscosity. The linear stability analysis is performed in chapter III to provide the parameter ranges where ripples can appear. The weak nonlinear analysis in chapter IV yields then the conditions for stable ripple formation and provides the correlation between incident beam angle and ripple orientation. Chapter V summarizes. In the appendix two models are presented for the coupling of external beams to gradients of the surface leading to surface currents.

\section{Liquid formulas}

\subsection{Evolution equations for velocity and height}

The following derivation follows closely the one found in text books, e.q. \cite{Beste,Light79,Pan13} with the additional consideration of viscosity and external currents. In order to see transparently which approximations are used we repeat the steps here.
\begin{figure}[]
\includegraphics[width=7cm]{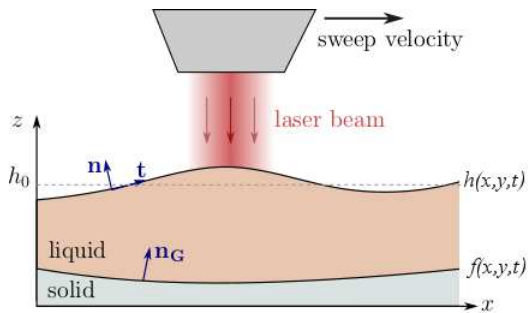}
\caption{Liquid layer with free surface $h(x,y,t)$ on a solid substrate exposed to an external laser or particle beam which creates a surface current due to a gradient of the surface. The vectors $\V n$ and $\V t$ describe the normal and tangent direction to the free surface. The bottom geometry $f(x,y,t)$ may be space and time dependent. \label{Fig:geometry} }
\end{figure}

\subsubsection{Bulk evolution equations}
We consider an incompressible fluid layer as depicted in Fig. \ref{Fig:geometry}. The motion of the viscous liquid is described by the Navier-Stokes equation
\be
\rho\left [\p{t}\V v+(\V v\cdot \nabla)\V v\right ]=-\nabla p+\eta\nabla^2 \V v+\V f 
\label{Navier}
\ee
where $\V v=(u,v,w)$ is the bulk velocity of the fluid, $\rho$ is the density, $\eta$ is the dynamic viscosity, $p$ is the pressure and $\V f$ is an external force. We assume $\V f=-\nabla U$ for a constant gravitation $U=U_0+\rho g z$ and fix the potential energy at $z=0$ with $U_0=-\rho g h_0-p_0$ with $h_0$ being the mean height on top of which the structure forms. 
Here, we consider a fluid layer with a free surface $h(x,y,t)$ on a solid substrate. The bottom geometry is described by $f(x,y,t)$ as depicted in Fig. \ref{Fig:geometry} and may in general vary in space and time.

We consider here flat weld pool shapes with a Maragoni number of about $200$ which is dimensions below the one for the onset of turbulence \cite{KKK16} and above the value of about $60$ that convection happens \cite{LCK00}. 
Therefore we assume an irrotational flow $\V v=\nabla \Phi$  and introduce the potential $\Phi$ which fulfills
\be 
\nabla^2\Phi=0
\label{place}
\ee
due to the incompressibility. The Navier Stokes equation (\ref{Navier}) can then be integrated once for an irrotational flow to yield
\be
\p{t}\Phi=-{1\over \rho}(p+U)-\frac 1 2 (\nabla \Phi)^2 \, .
\label{Euler}
\ee
 The term due to the viscosity vanishes in the bulk liquid for the  assumed incompressible fluid (\ref{place}).

\subsubsection{Boundary conditions at top and bottom}

Next, we consider the boundaries at the top and bottom of the liquid layer:
At the bottom $z=f(x,y,t)$, the velocity component normal to the interface corresponds to the temporal change of the bottom topology
\be
\nabla \Phi \cdot \V n_G = \partial_t f \quad \text{at} \quad  z=f(x,y,t)
\label{bc:bottom}
\ee
where the normal vector of the ground is given by $\V n_G=(-\nabla_2 f,1)/\sqrt{1+(\nabla_2f)^2}\approx (-\nabla_2 f,1)$ with $\nabla_2=\partial_x^2+\partial_y^2$.
At the free surface $z=h(x,t)$, the force equilibrium
\be
\Pi\cdot \V n=(\gamma \nabla_2 h) \V n  + (\V t\cdot \nabla \gamma) \V t \quad \text{at} \quad  z=h(x,y,t)
\label{bc:forceequilibrium}
\ee
holds, where
\be
\Pi_{ij}=-p \delta_{ij}+\eta (\p{i}v_j+\p{j}v_i) 
\ee 
is the stress tensor for a viscous, incompressible fluid and $\V n$ and $\V t$ are the normal and tangential vectors of the free surface, respectively \cite{KT07}. In the following we abbreviate $h_x=\partial_x h$ for legibility.
 Using the normal vector $\V n=(-h_x,-h_y,1)/\sqrt{1+h_x^2+h_y^2}\approx (-h_x,-h_y,1)$ 
and projecting Eq. (\ref{bc:forceequilibrium}) onto $\V n$, we find
\ba
 &p -p_a=-\gamma \nabla_2^2 h +2\eta \biggl [h_x^2\Phi_{xx}+h_y^2\Phi_{yy}+\Phi_{zz}\nonumber\\ 
 & +2(h_x h_y\Phi_{xy} -h_x\Phi_{xz}-h_y\Phi_{yz})\biggr ]
 \label{bc:normal}
\end{align}
at $z=h(x,y,t)$. Here we have assumed $h_x^2\ll 1$ and $h_y^2\ll 1$ to neglect the denominator of the inverse curvature radius. 
In addition to the well-known Laplace pressure contribution, the pressure contains terms due to viscosity.
Projecting the force equilibrium (\ref{bc:forceequilibrium}) onto the tangential vector  $\V t=(1,0,h_x)/\sqrt{1+h_x^2}\approx (1,0,h_x)$ yields
\be
(\p{x}+h_x \p{z} )\gamma&=&2\eta [-h_x^2\Phi_{xz}-h_xh_y\Phi_{yz}-h_y\Phi_{xy} \label{bc:tangential}
\nonumber\\&&+h_x(\Phi_{zz}-\Phi_{xx})+\Phi_{xz}] 
\ee
at $z=h(x,y,t)$ where we chose $\V t$ in $x$ direction without loss of generality. 

Introducing Eq. (\ref{bc:normal}) for the pressure into the bulk equation for potential flow (\ref{Euler}) at the free surface $z=h(x,y,t)$, we get
\ba
&\p{t}\Phi=-g(h-h_0)-\frac 1 2 (\nabla \Phi)^2+{\gamma\over \rho} \nabla_2^2h \label{EulerAth}
\\&
\!-\!2\eta \left [h_x^2\Phi_{xx}\!+\!h_y^2\Phi_{yy}\!+\!\Phi_{zz} 
\!+\!2(h_x h_y\Phi_{xy}\!-\!h_x\Phi_{xz}\!-\!h_y\Phi_{yz})\right ] 
\end{align}
at $z=h(x,y,t)$ where $p_a=p_0(z=h)$.

The velocity field at the free surface $h(x,y,t)$  is connected to the temporal change of this interface via a kinematic boundary condition
\ba
&\p{t} h=w -u h_x -v h_y+D_x(\theta)h_{xx}+D_y(\theta)h_{yy} 
\label{kin}
\end{align}
at $z=h(x,y,t)$. This contains the velocity $w$ in $z-$direction and the projection of the horizontal velocity to the normal vector $\V n$.
Since the laser intensity can change the melt morphologies \cite{PDS87}, we consider the induced surface current due to the coupling of an external beam to the surface gradient 
\be
J_x&=&-D_x(\theta) h_x,\quad
J_y=-D_y(\theta) h_y
\ee
derived in appendix~\ref{curr} which is dependent on the incident angle $\theta$ of impact to the surface. This surface current is coupled here in a conserving way to the change of height $\partial_t h=...-\nabla\cdot \V J=...+D_x h_{xx}+D_yh_{yy}$.

\subsubsection{Rescaling}

\begin{figure}[]
\includegraphics[width=6cm]{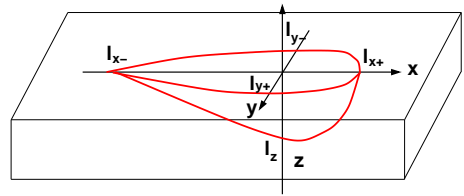}
\caption{\label{shape} 
Sketch of the form of liquid pot due to laser melting sweep. The dimensions are $l_1=l_{y-}\approx l_{y+}\approx l_{x+}\approx l_z \approx 0.1-10$mm and $l_{x-}\approx 3-10\times l_1$.}
\end{figure}

Three-dimensional finite-element
models have been successfully developed to predict the laser welding modes \cite{BSSS09}. These experiments suggest a form of the laser-induced liquid pot as illustrated in figure~\ref{shape}. The geometry is nearly symmetric with the dependence of the typical size 
\be
l\sim \sqrt{\kappa {d\over v}}
\ee
on the thermal conductivity $\kappa$, 
the spot-size $d$ and the sweep velocity $v$. The elongation is dependent on the timescale of cooling and freezing which is in $ms$ range. As illustrated in figure~\ref{shape} the length $l=l_{x+}+l_{x-}$ and $h_0=l_z$ are the characteristic melt pool length and depth respectively. They depend on the laser intensity and the scanning speed, so they can be easily varied in the experiments, but it is difficult to measure them precisely. The estimated values are $l\sim 1-10$\,mm and $h_0\sim 0.2-2$\,mm for conductive laser welding and for LIPSS one has $l\sim 10^{-2}-10^{-1}$\,mm, $h_0\sim 1-5\times10^{-4}$\,mm. The $l/h_0$-ratio depending on the laser scanning speed is in the range $l/h_0\sim3-10$ for conductive laser welding and $l/h_0\sim10^{2}$ for fs-LIPSS. 

It is now convenient to use dimensionless values by introducing the scaling
\ba
&\begin{pmatrix}x\cr y \end{pmatrix}\!\to\! \begin{pmatrix}x\cr y \end{pmatrix} l, \quad  \begin{pmatrix}h\cr f \end{pmatrix}\!\to\! \begin{pmatrix}h\cr f \end{pmatrix} h_0,\,  \quad z\!\to\! z h_0,\,
\nonumber\\
&t\!\to\! t \tau, \quad \Phi\!\to\!\Phi {l^2\over \tau}, \quad D\!\to\!D {l^2\over \tau}
\label{scal}
\end{align}
with some characteristic time scale $\tau$. For laser processing with pulsed lasers, this time scale (the period of the excitation) can be assigned to the laser repetition rate if the surface does not solidify in the time interval between the pulses \cite{OPIKRYEI13,CLSFS14}. Femtosecond laser pulses melt the surface only for a time interval of $\lesssim 10^{-9}$\,s, which is shorter than the inter-pulse delay of the majority of available lasers, but in this case the periodic excitation may come from the interference between the incident and the surface-scattered waves \cite{GLBN20}. If the beam of a continuous wave (cw) laser is scanned over the sample, as it is done e.g., by laser welding, the excitation of each point at the surface changes with time and can be Fourier-transformed to a broad band of frequencies. If the sweep or scanning velocity is accordingly tuned the surface instabilities can freeze such that an instant picture of the surface ripples is taken.
An overview about possible instabilities depending on the welding speed and current can be found in \cite{Wei12}.

 \begin{table}
 \begin{tabular}{|c|c|c|c|c|c|}
\hline
 & $\gamma$ 
& $\rho$ & $\eta$ 
& $l$ & $h_0$
\\
\hline
 & $\dfrac{N}{m}$ 
& $10^{4}\dfrac{kg}{m^3}$ & $10^{-3}Pa\!\cdot \!s$ 
& $ 10^{-3}m $ & $ 10^{-3}m$
\\&
&&
&&\\
\hline 
Au & $1.1$ 
& $1.7$& $4$  
& $1(1-10)$ & $1(0.2-2)$\\
Fe & $1.8$ 
& $0.7$& $6$ 
& $1 (1-10)$ & $1 (0.2-2)$\\
\hline 
\end{tabular}
\caption{
Materials parameters of liquid gold and iron. Here $\gamma$ is the surface tension, 
$\rho$ is the density, and
$\eta$  is the melt dynamic viscosity.
For the length and height $l=h_0=1\times 10^{-3}$m is chosen though it varies in a certain range typical for conductive laser welding.}
 \label{TableParam}
\end{table}

We introduce the four dimensionless parameters
\ba
G&=&{g h_0\tau^2\over l^2},\quad \Gamma ={\gamma\tau^2 h_0\over \rho l^4},\quad H={\eta \tau \over \rho l^2},\quad \delta={h_0\over l} \, .
\label{para}
\end{align}
for the gravitation $G$, viscosity $H$, and surface tension $\Gamma$ as well as expansion parameter $\delta$ with the values for $Au$ and $Fe$ seen in table~\ref{TableParam}. Up to this point the model is universal in the sense, that it is applicable to all types of periodic laser induced structures (including fs-LIPSS), in which formation of the liquid phase on the surface is involved. Now we have to focus on some particular range of the parameters to be able to  estimate the influence of different terms in the Navier-Stokes equation. Most of practically important processes like laser welding or cutting provide geometrical melt dimensions listed in the table~\ref{TableParam}. For LIPSS formation processes other than purely hydrodynamical effects have to be taken into account for ripple formation.
Thus, we obtain
\ba
&Au:\nonumber\\
&G=10^4{\tau^2\over s^2}; \Gamma =6.5 \times 10^4 {\tau^2\over s^2}; H=0.24 {\tau\over s}, \delta =0.002-2 \nonumber\\
&Fe:\nonumber\\
&G=10^4{\tau^2\over s^2}; \Gamma =26 \times 10^4 {\tau^2\over s^2}; H=0.86 {\tau\over s}, \delta =0.002-2,
\label{par}
\end{align}
respectively. Using as typical time scale a value of $\tau=10^{-2}s $ such that $G=1$ we have
\be
&&Au:\quad \Gamma=6.5,\quad H=2.4\times 10^{-3}
\nonumber\\
&&Fe:\quad \Gamma=26,\quad H=8.7\times 10^{-3}
\label{typ}
\ee
which show for both cases that the viscosity parameter is small compared to the surface tension. Let us note that with the characteristic time we have the freedom to chose also another scaling estimating $\tau$ e.q. by the squared beam diameter divided by thermal diffusion leading to $\tau \!\approx\!\ 3\!\times\! 10^{-5} s$. Then the parameters would take the values
\ba
&Au:\quad 
G\!=\!0.9\!\times\!10^{-5}, \,
\Gamma\!=\!5.8\!\times\! 10^{-5},\,
H\!=\!0.7\!\times\! 10^{-5}
\nonumber\\
&Fe:\quad 
G\!=\!0.9\!\times\!10^{-5}, \,
\Gamma\!=\!23\!\times\! 10^{-5},\,
H\!=\!2.6\!\times\! 10^{-5}.
\label{typ1}
\end{align}
We will work with the values (\ref{typ}) and will discuss if the results are dependent on the choice of $\tau$.

With (\ref{scal}) the incompressibility condition (\ref{place}) for the velocity potential is expressed as
\be
(\p{x}^2+\p{y}^2)\Phi+{1\over \delta^2}\p{z}^2\Phi=0 .
\label{place1}
\ee
The kinematic boundary condition (\ref{kin}) in dimensionless coordinates reads 
\be
\p{t} h\!-\!\left .{1\over \delta^2}\p{z}\Phi\right. =-\nabla_2h\cdot \nabla_2\Phi\!+\!\left (D_x\p{x}^2\!+\!D_y\p{y}^2\right ) h
\label{kin1}
\ee
at the free interface $z=h(x,y,t)$
and the boundary condition (\ref{bc:bottom}) at the bottom $z=f(x,y,t)$ becomes
\be
\p{t} f&=-\nabla_2\Phi\nabla_2 f+{1\over \delta^2}\p{z} \Phi.
\label{ground}
\ee
The other boundary conditions Eq. (\ref{EulerAth}) and (\ref{bc:tangential}) at $z=h(x,y)$  take the form
\ba
&\p{t}\Phi=-G(h\!-\!1)+\Gamma \nabla_2^2h-{(\nabla_2\Phi)^2\over 2}-{(\p{z}\Phi)^2\over 2\delta^2}
\nonumber\\&
-2H \biggl[h_x^2\Phi_{xx}+h_y^2\Phi_{yy}+{\Phi_{zz}\over \delta^2}
\nonumber\\&
\qquad\qquad+2\left (h_x h_y\Phi_{xy}-{1\over \delta}h_x\Phi_{xz}-{1\over \delta}h_y\Phi_{yz}\right )\biggr ]
\label{vel}
\end{align}
and 
\be
\left (\p{x}+{h_x\over \delta} \p{z} \right )\Gamma&
= &2\delta H \biggl [{-h_x^2\Phi_{xz}-h_xh_y\Phi_{yz}+\Phi_{xz}\over \delta}
\nonumber\\&&-h_y\Phi_{xy}+{h_x\over \delta^2}\Phi_{zz}-h_x\Phi_{xx}\biggr ] \, .
\label{17}
\ee
These equations (\ref{place1}) - (\ref{17})   form a closed system.

\subsection{Shallow water approximation}

Though the parameter $\delta$ varies for the experimental spots between $0.02-2$ 
according to table~\ref{TableParam} we will employ the idea of shallow-water approximation to consider the parameter $\delta\ll1$. 
We expand then the potential
\be
\Phi=\Phi_0+\delta^2\Phi_1
\ee
and get from (\ref{place1})
\be
\p{z}^2\Phi_0=0,\quad \p{z}^2\Phi_1=-\nabla_2^2\Phi_0.
\label{Phi0Phi1}
\ee
Simple integration of the first equation introduced in the second one provides
\ba
\Phi_0&=\Phi_{00}(x,y,t)+z\Phi_{01}(x,y,t)
\nonumber\\
\Phi_1&=-{z^3\over 6} \nabla_2^2\Phi_{01}-{z^2\over 2}\nabla_2^2\Phi_{00}
+z c_1(x,y,t)+c_0(x,y,t).
\label{phi1}
\end{align}
We use this form in the condition for the ground (\ref{ground}) which leads to
\ba
\p{t} f\!+\!\nabla_2\Phi_0\!\nabla_2 f\!+\!{f^2\over 2}\!\nabla_2^2\Phi_{01}\!+\!f\nabla_2^2\Phi_{00}\!-\!c_1\!-\!{\Phi_{01}\over \delta^2}\!=\!o(\delta^2)
\end{align} 
providing $\Phi_{01}=0$ and the function $c_1$. This determines the $z$-dependence of the velocity potential (\ref{phi1})
\ba
\Phi_0&=\Phi_{00}(x,y,t)
\nonumber\\
\Phi_1&=\left (z f-{z^2\over 2}\right )\nabla_2^2\Phi_{00}
+z \left (\p{t} f+\nabla_2 \Phi_{00}\nabla_2 f \right )+c_0
\label{phi2}
\end{align}
and
the kinematical boundary condition (\ref{kin1}) becomes 
\be
\p{t}(h\!-\!f)&=&\left [(f\!-\!z) \nabla_2^2\Phi_{00}\!-\!\nabla_2(h-f)\cdot \nabla_2 \Phi_{00}\right ]_{z=h}
\nonumber\\
&&\!+\!\left (D_x\p{x}^2\!+\!D_y\p{y}^2\right ) h
\nonumber\\
&=&\nabla_2\left [(f\!-\!h)\nabla_2 \Phi_{00}\right ]\!+\!\left (D_x\p{x}^2\!+\!D_y\p{y}^2\right ) h.
\label{23}
\ee

Multiplying the Euler equation (\ref{vel}) with $\delta^2$ and using (\ref{phi2}) one obtains up to $o(\delta^3)$
\ba
&
\p{t} \Phi_{00}=-G(h-1)+\Gamma \nabla_2^2h -\frac 1 2 (\nabla_2 \Phi_{00})^2
\nonumber\\
&-2 H\left [
(h_x^2-1)\p{x}^2+(h_y^2-1)\p{y}^2+2 h_x h_y \p{xy}^2 
\right ]\Phi_{00}.
\end{align}
Remembering again $h_x^2\ll 1$ and $h_y^2\ll 1$ we obtain together with
the kinematical boundary condition (\ref{23}) the final coupled equation system
\be
\p{t}\Phi_{00}&=&-G(h-1)+\Gamma \nabla_2^2 h+2 H\nabla_2^2\Phi_{00}-\frac 1 2 (\nabla_2 \Phi_{00})^2
\nonumber\\
\p{t} h&=&\nabla_2[(f\!-\!h)\nabla_2\Phi_{00}]\!+\!\p{t} f\!+\!\left (D_x\p{x}^2\!+\!D_y\p{y}^2\right ) h\nonumber\\
&&
\label{sys}
\ee
which allows to determine the surface profile $h(x,y,t)$ and the velocity potential $\Phi_{00}$ in dependence on the three parameters gravitation $G$, viscosity $H$, and surface tension $\Gamma$. These equations in lowest order $\delta^2$ correspond to the shallow water equations with time-dependent bottom and surface tension \cite{Beste}. Similar expansions have been performed in \cite{Dias08,MNMM21}. 

From the transverse boundary condition (\ref{17}) we obtain
\be
(\delta \p{x}+h_x\p{z})\Gamma=+o(\delta^2)
\ee
as condition for a possible spatial dependence of the surface tension $\Gamma$.
In the following, we neglect gradients in the surface tension  which can e.g. occur due to gradients in temperature or chemical gradients along the interface and use $(\p{x}+h_x \p{z} )\gamma=0$.

\subsection{Formulation in terms of the velocity}
When we introduce a velocity field connected to the gradient of the potential
\be
\V u=\begin{pmatrix} u \cr v \end{pmatrix} = \nabla_2 \Phi_{0} \, ,
\label{eq:u0v0}
\ee
we can reformulate the shallow water equations (\ref{sys}) as
\be
\partial_t h =& -\partial_x [(h-f) u] - \partial_y [(h-f) v] + \partial_t f\nonumber\\
&+\left (D_x\p{x}^2+D_y\p{y}^2\right ) h\nonumber\\
\partial_t u 	=& -G \partial_x h + \Gamma \partial_x \nabla_2^2 h + 2H \partial_x (\partial_x u + \partial_y v)\nonumber\\
 				& - u (\partial_x u)- v (\partial_x v)\nonumber\\
\partial_t v 	=& -G \partial_y h + \Gamma \partial_y \nabla_2^2 h + 2H \partial_y (\partial_x u + \partial_y v) \nonumber\\
& - u (\partial_y u)- v (\partial_y v) \, .
\ee
Due to Eq. (\ref{eq:u0v0}), the relation $\partial_x v = \partial_y u$ holds, so that we can also write 
\ba
&\partial_t (h-f) = -\nabla_2 \cdot [(h-f) \V u] + \left (D_x\p{x}^2+D_y\p{y}^2\right ) h
\label{vel2a}\\
&\left (\partial_t\!+\!\V u\cdot \nabla_2\right ) \V u=-\nabla_2( G h \!+\! \Gamma  \nabla_2^2 h \!+\! 2H \nabla_2\cdot \V u)
\label{vel2}
\end{align}
where we have used $\nabla_2^2\V u=\nabla_2 (\nabla_2\cdot \V u)$ due to the curl-free condition (\ref{eq:u0v0}).
The latter equation shows how the Navier-Stokes equation (\ref{Navier}) has translated into the coupled equations for the two-dimensional velocity and the height. Especially the right-hand side of (\ref{vel2}) shows how the pressure gradient, viscosity and gravitational forces combine.

It is important to note here that the two-component velocity $\V u$ is not two-dimensional divergence-free, i.e. $\nabla_2\cdot \V u\ne 0$, compared to the three-dimensional velocity $\V v$ which is divergence-free due to incompressible fluid. Therefore the viscosity term in the Navier Stokes equation (\ref{Navier}) vanishes but reenters the theory by the surface condition (\ref{bc:normal}).

\subsection{Conservation laws}

It is instructive to analyze the conservation laws. From the right-hand side of (\ref{vel2}) it is visible that the total matter is conserved
\ba
&\p{t}\int \!\!\!d^2 r [h(x,y,t)-f(x,y,t)]
\nonumber\\
&=\iint  \!\!dx dy \left \{
\nabla_2[(f-h) \V u]\!+\!\left (D_x\p{x}^2\!+\!D_y\p{y}^2\right ) h\right \}
\nonumber\\
&=\left .(f-h) \V u \right |_{\partial}  
\!+\!\left .\left (D_x\p{x}\int \!\!dy\!+\!D_y\p{y}\int \!\! d x\right ) h\right |_{\partial}=0
\end{align}
if we demand that
\be
\left .\V u\right |_{\partial}=\left .\nabla_2\Phi_{00}\right |_{\partial}=0,\qquad\left .\nabla_2h\right |_{\partial}=0
\ee
at the boundaries.

The momentum balance is derived using (\ref{vel2a}) and (\ref{vel2}) in appendix~\ref{mbalance} to obtain
\be
\p{t}[(h-f)u_i]=-\p{j} \Pi_{ij}-\p{i} V+s_i
\label{pbal}
\ee
with the effective momentum current density
\be
\Pi_{ij}=(h-f) u_iu_j+2 H (h-f) \p{i}u_j
\label{stress}
\ee
which shows that the viscosity enters if the velocity has a spatial variation. The potential becomes
\be
V=G{h^2\over 2}+\Gamma \left [h\nabla_2^2 h-{(\nabla_2h)^2\over 2}\right ]
\label{V}
\ee
where one sees the contribution of the surface tension besides the gravitational potential. The remaining term in (\ref{pbal}) reads
\be
\V s=-2 H \p{j}(h-f)\p{j}\V u+f\nabla (G h+\Gamma \nabla^2h)
\label{s}
\ee
and acts as a source when integrating (\ref{pbal})
\ba
&\p{t}\int\!\!\!d^2r(h-f) \V u
\nonumber\\
&=\int\!\!\!d^2r\,\V s=\int\!\!\!d^2r\left [2 H \p{j}(f\!-\!h)\p{j}\V u
\!-\!(G h\!+\!\Gamma \nabla_2^2 h)\nabla_2 f\right ].
\end{align}
We see that the spatial-dependent ground has an impact on the momentum balance by coupling to gravitation and surface tension. The viscosity  couples again with the spatial dependence of the velocity and the ground. 

The effect of viscosity can be rewritten from a source or damping term into a modification of the mean momentum velocity. In fact we can rewrite (\ref{pbal}) alternatively into
\be
&&\left \{\p{t}+[u_j-2 H\p{j}\ln(h-f)] \p{j}\right \} [(h-f)u_i]
\nonumber\\
&&=-\p{j} \bar \Pi_{ij}-\p{i} V+\bar s_i
\label{pbal1}
\ee
where the momentum current density $\bar \Pi$  contains only the viscosity part of (\ref{stress}) and the velocity gradient appears instead of the velocity in the source term
\ba
\bar {\V s}=f\nabla (G h\!+\!\Gamma \nabla^2h)\!-\!\V u  \left [(h\!-\!f)\p{j} u_j\!+\!2 H {(\p{j} (h\!-\!f))^2\over h\!-\!f}\right ].
\label{s1}
\end{align}

We can summarize that the approximate equations derived from Navier-Stokes equation with respect to the surface and the two-dimensional velocity obeys conservation laws for mass (volume) and momentum. We can identify the gravitational potential and the potential by the surface tension (\ref{V}). The effect of viscosity is that it modifies the stress tensor (\ref{stress}) and the damping (\ref{s}) or alternatively changes the mean velocity of momentum by effectively
\be
\V u\to \V u-2 H\nabla_2\ln (h-f)
\ee
visible from the substantial derivative in (\ref{pbal1}). 

The bottom effectively induces a source of momentum transfer if it has a nonzero spatial gradient. This momentum transfer appears by gravitation and surface tension. The surprising coupling of the latter one appears together with the second derivative of the surface.

\section{Linear stability analysis}

\subsection{Constant external current}

We linearize the system (\ref{sys}) with respect to small time- and space-dependent perturbations
\ba
\Phi_{00}(\V r,t)&=\bar \Phi_{00}+\delta \Phi(\V r,t)\nonumber\\
h(\V r,t)&=h_0+\delta h(\V r,t)
\nonumber\\
f(\V r,t)&=\delta f(\V r,t)\nonumber\\
D_x\partial_x^2h+D_y\partial_y^2 h&=(D_x\partial_x^2+D_y\partial_y^2)\delta h(\V r,t)
\label{lin}
\end{align}
where $h_0=1$ due to the scaling (\ref{scal}) and the time is in units of $\tau$ and space in units of $l$.  The bottom $\delta f(\V r,t)$ and the induced surface current contribution represented by the $D$ are the sources of disturbance which will provoke a $\delta h$ and $\delta \Phi$. Due to the second derivatives only the terms $D\partial^2\delta h(r,t)$ contributes to the linear response and any time dependence of $D$ is considered in chapter~\ref{Floquet}.

First we consider a constant external current $D(t)=D$. 
Linearizing (\ref{sys}) by (\ref{lin}) and after Fourier transform ${\rm e}^{-i\omega t+i \V k\V r}$  of time and space one gets
\ba
&(-i \omega +2 H k^2)\delta \Phi+(G+k^2 \Gamma)\delta h=0\nonumber\\
&-k^2\delta \Phi+(-i \omega+D k^2) \delta h=-i\omega \delta f
\label{disp}
\end{align}
dependent on the direction due to the external diffusion current
\be
D=D_x(\theta){k_x^2\over k^2}+D_y(\theta) {k_y^2\over k^2}.
\ee
We are searching for the eigenmodes of the systems (\ref{disp}) which means to consider (\ref{disp}) $\delta f=0$. 
The different regimes of instability can be best discussed by the growth rate $\lambda=-i\omega$. One obtains
\be
\lambda &=&\frac 1 2 \left (-{b}\pm\sqrt{b^2-4 a}\right )\nonumber\\
a&=&G k^2+(2 D H+\Gamma)k^4\nonumber\\
b&=&(2 H+D) k^2.
\label{af}
\ee

\begin{figure}[h]
\includegraphics[width=9cm]{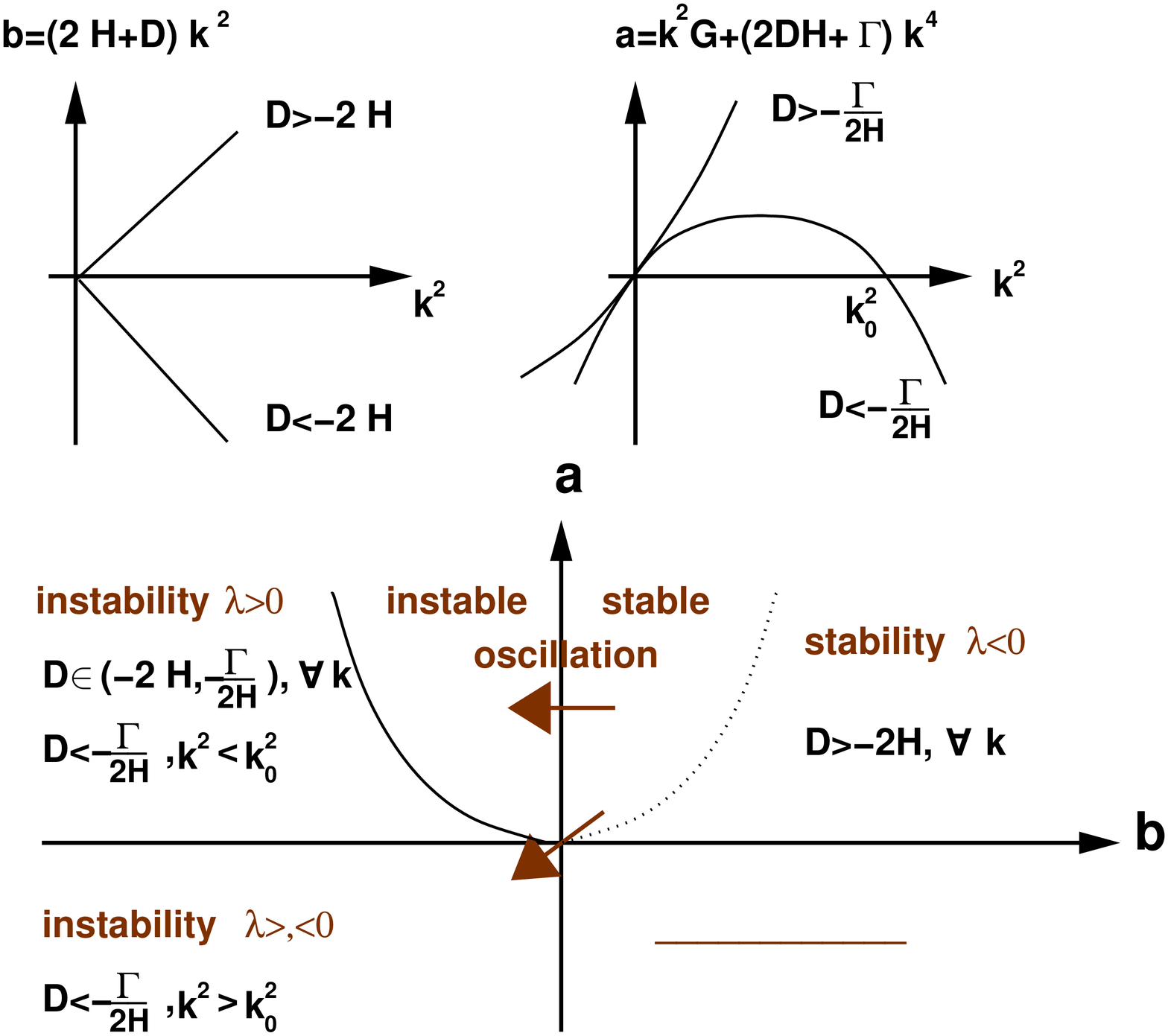}
\caption{\label{stability} 
The wavelength dependence of $b(k)$ and $a(k)$ (above) determining the growth rate  
(\ref{af}) which allows to discuss four different regions (below) according to $a\gtrless0$ and $b\gtrless 0$. Here 
$k_0^2=-G/(2DH+\Gamma)$ and all indicated relations are holding for both scalings of $\tau$ in (\ref{typ}) or (\ref{typ1}).} 
\end{figure}

For a given wave number $k$, a positive (negative) $\lambda$ value indicates that this mode is unstable (stable) and will grow (be suppressed) in amplitude. We proceed first with the stability analysis \cite{Cross:2009,AM12} illustrated in Fig.~\ref{stability} . The system is stable for $a>0$ and $b>0$ since then $\lambda<0$ for all $k$ and it has an oscillatory solutions if $b^2<4 a$.  The sign change of $a$ and $b$ in dependence on the wave vector can be seen in the upper figure \ref{stability}. It depends on the relative values of the diffusion coefficients $D$ of the external current and combinations of the viscosity $H$, surface tension $\Gamma$ and gravitational constant $G$. 

In order to discuss these different regions more in detail we observe that for our parameters (\ref{typ}) it holds
\be
-{\Gamma\over 2 H}<2 (H-\Gamma)&<&2(H-\sqrt{\Gamma})<-2 H
\nonumber\\&<&2(H+\sqrt{\Gamma})<2 (H+\Gamma)
\label{bed}
\ee
and for a different timescale (\ref{typ1}) the second with the third and the six with the seventh terms in (\ref{bed}) have to be interchanged. The different regimes can be derived then straightforwardly as illustrated in the lower figure~\ref{stability}.

Within the continuous change of the parameters we can reach the two adjacent instability regions from the stable one by the two arrows indicated in Fig.~\ref{stability}. The right lower quarter is not possible with our parameters (\ref{bed}).  We are left with two different possible paths from stability to instability: (i) 
$b<0$ and $a<b^2/4$
for stationary-growing patterns and (ii) $b<0$ and $a > b^2/4 >0$ for oscillatory patterns since the square-root term becomes purely imaginary for the growth rate $\lambda$ in (\ref{af}). We see that instability is only possible for $b<0$ for our parameter sets. Since $H$ is always positive, this means that $D$ needs to be negative for instability. The externally induced current then acts like a diffusion term with negative diffusion coefficient, thereby provoking a roughening of the  surface.

\begin{figure}[h]
\includegraphics[width=4cm]{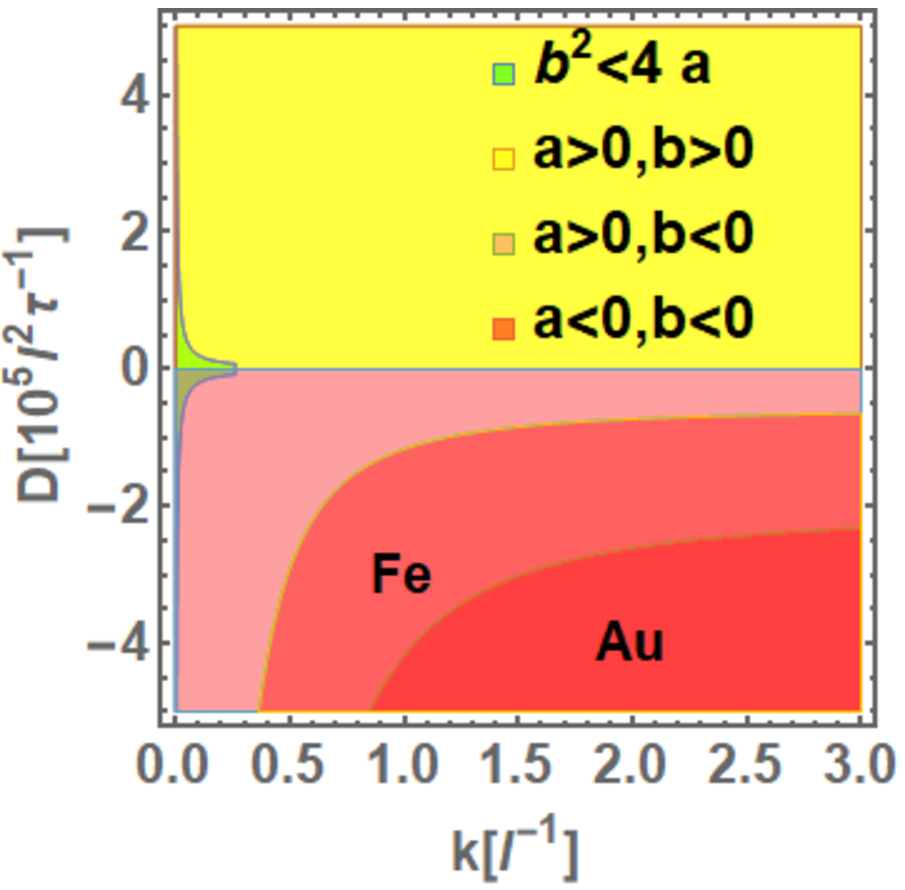}\includegraphics[width=4cm]{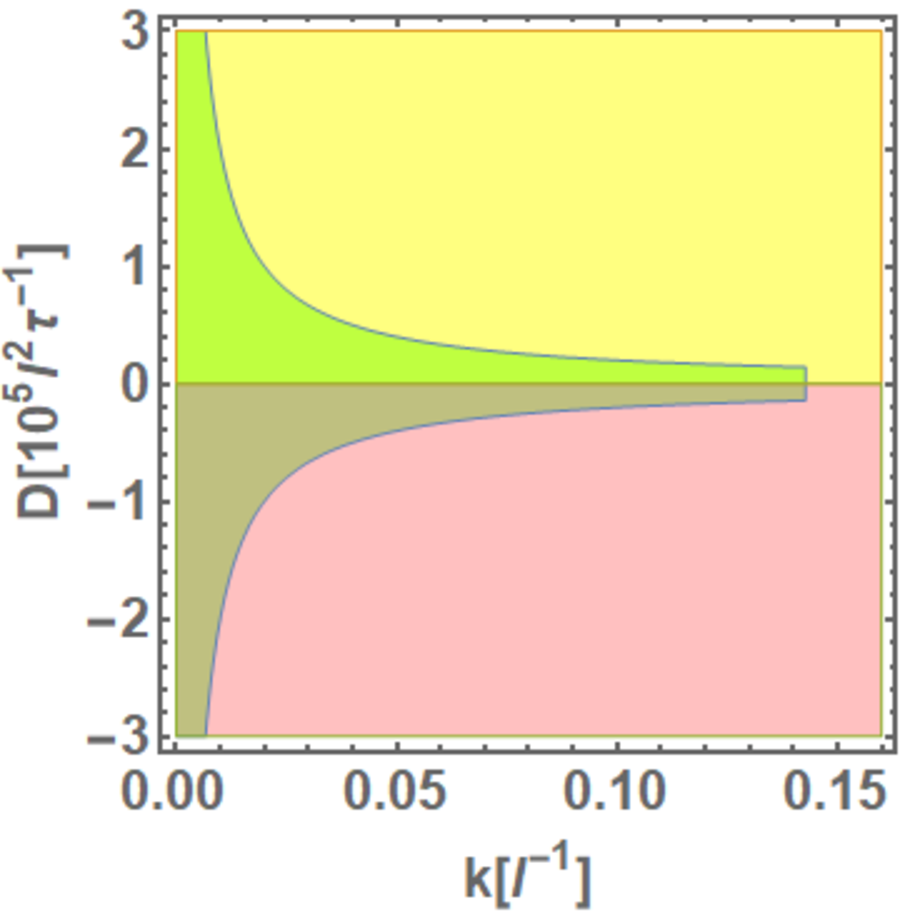}
\caption{\label{regions} The areas of stable (yellow) oscillating (green) and unstable (red,pink) regions according to figure~\ref{stability} in dependence on the dimensionless wavelength and external current with parameters of (\ref{par}). Right side is a zoom of left figure. The data for $Fe$ and $Au$ differ visibly only for the range $a<0,b<0$}.
\end{figure}

In figure~\ref{regions} we present these regions in terms of wavelength and external current for the parameters (\ref{par}) of $Au$ which are qualitatively similar to $Fe$ except that the region $a<0,b<0$ differs. The plot is independent of the time scale $\tau$. One sees that the two unstable regions are appearing in separated regions of $D<0$. The exponential growing range appears only for larger wavevectors $k(D)$.
The oscillating behaviour appears for smaller wavevectors with an upper limiting value. Let us discuss this case more in detail. The major driving is the external current which dominates the difference between $Au$ und $Fe$ parameters. 

\begin{figure}[h]
\includegraphics[width=4cm]{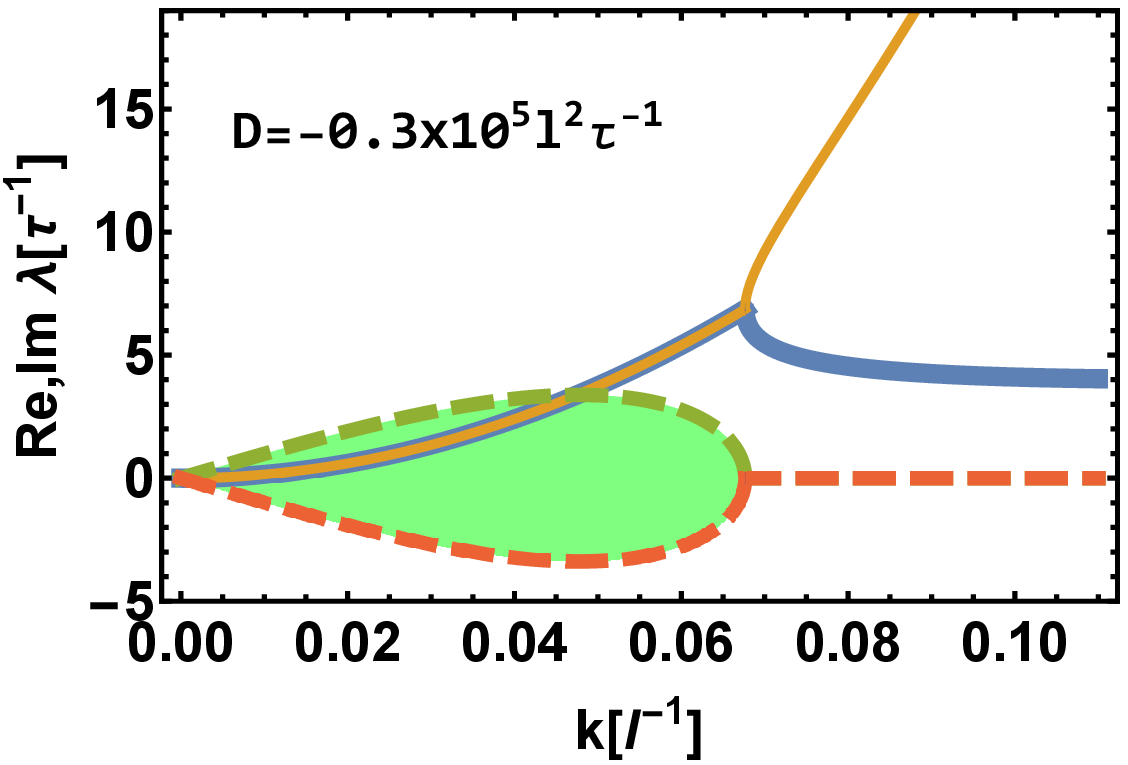}\includegraphics[width=4cm]{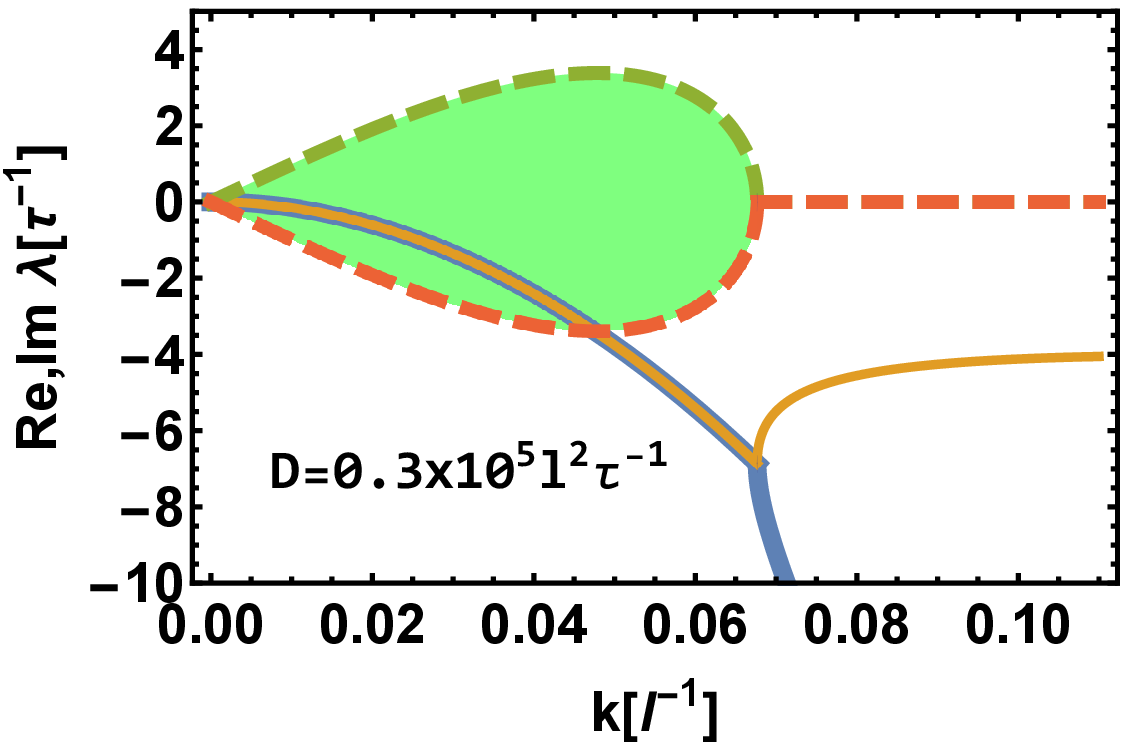}
\caption{\label{lamD} The real (solid) and imaginary (dashed) part of the growth rate $\lambda=-i\omega$ for a  horizontal cut of figure~\ref{regions} corresponding to unstable (left) and stable (right) behaviour. The oscillating range is indicated by filling.}
\end{figure}

In figure~\ref{lamD} we give the momentum dependence of the growth rate for an unstable $D<0$ and stable $D>0$ solution resulting in positive/negative real parts of the growth rates respectively. The oscillating area are indicated by the shading seen as finite imaginary part in the growth rate. The real part (solid line) shows a bifurcation at the wavevectors where the oscillation disappears. In principle larger wavevectors (smaller wavelength) show a faster growth rate than the oscillating modes in linear response such that in the unstable regime the exponential growing modes for smaller wavelength will win. 

\begin{figure}[h]
\parbox[]{8.7cm}{
\parbox[]{4.cm}{
\parbox[]{3.5cm}{
\includegraphics[width=3.5cm]{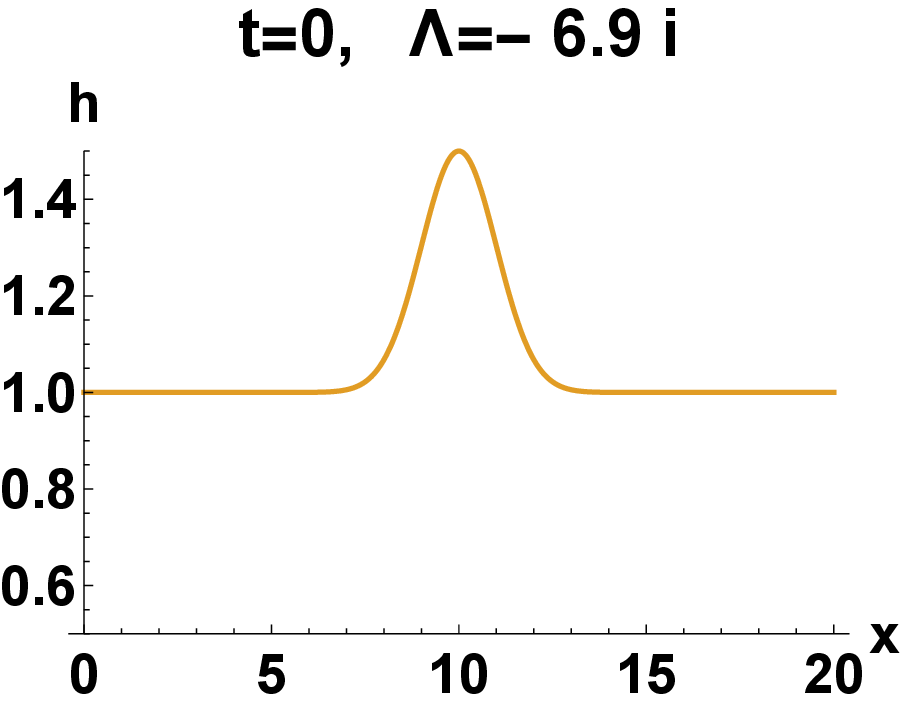}
\includegraphics[width=3.5cm]{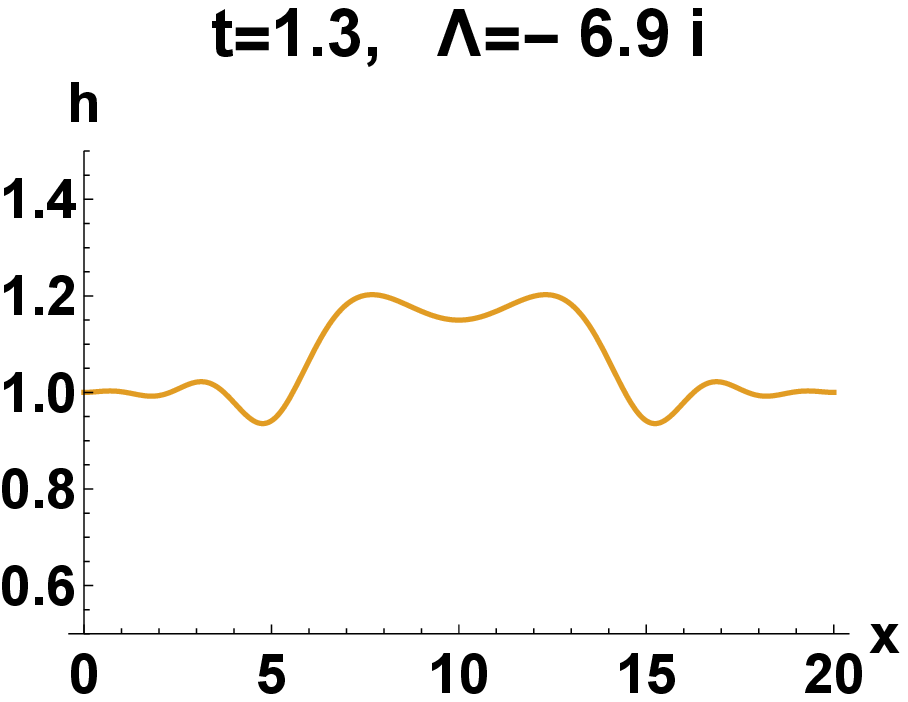}
\includegraphics[width=3.5cm]{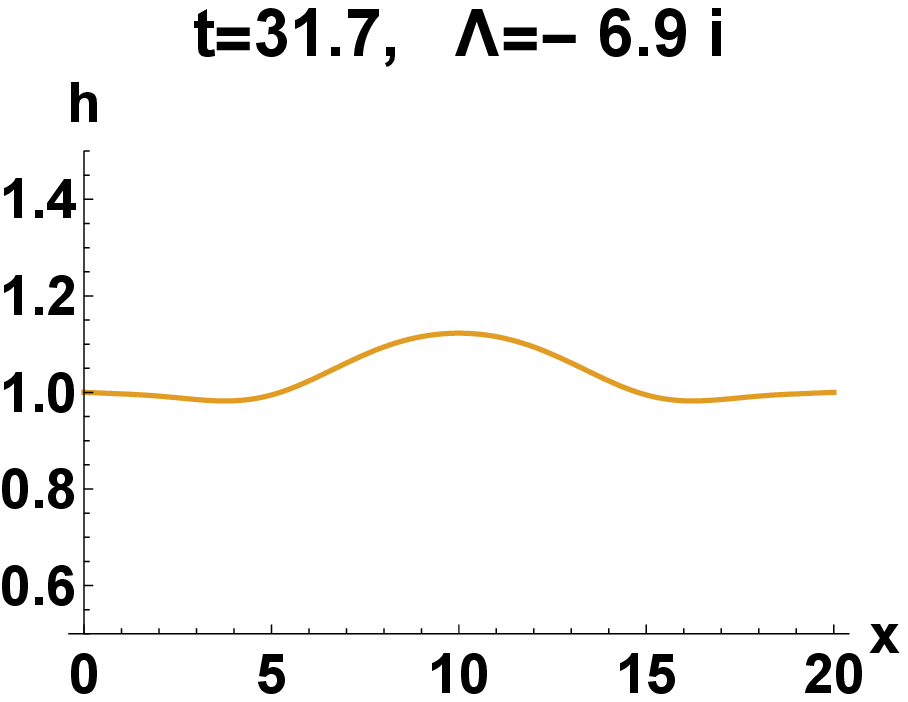}
}
\hspace*{-0.2cm}\parbox[]{0.3cm}{(a)\\[2cm](b)\\[2cm](c)}
}
\hspace{0.5cm}
\parbox[]{4.cm}{
\parbox[]{3.5cm}{
\includegraphics[width=3.5cm]{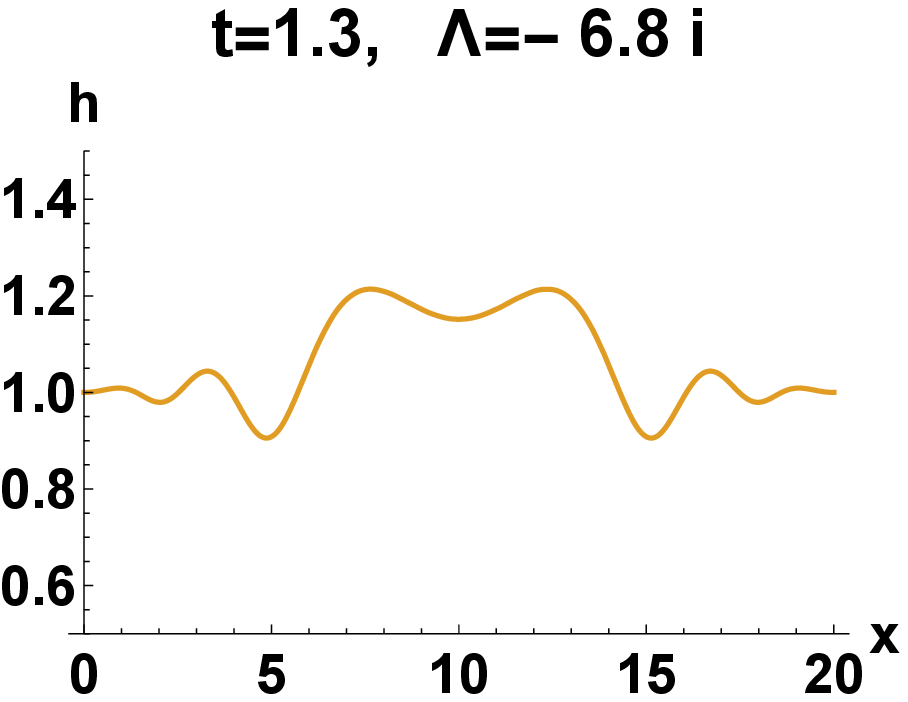}
\includegraphics[width=3.5cm]{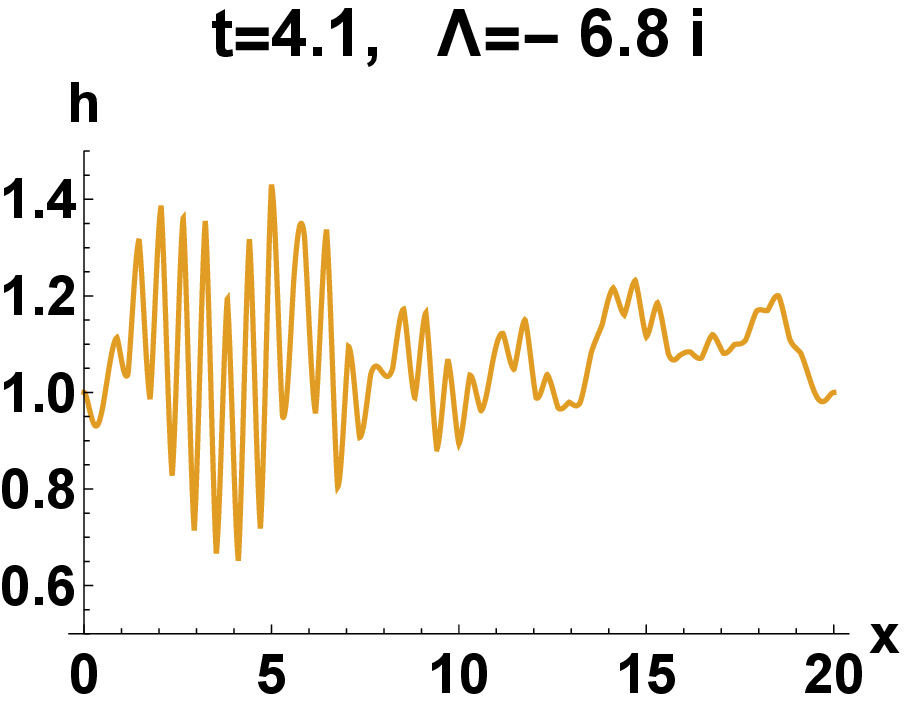}
\includegraphics[width=3.5cm]{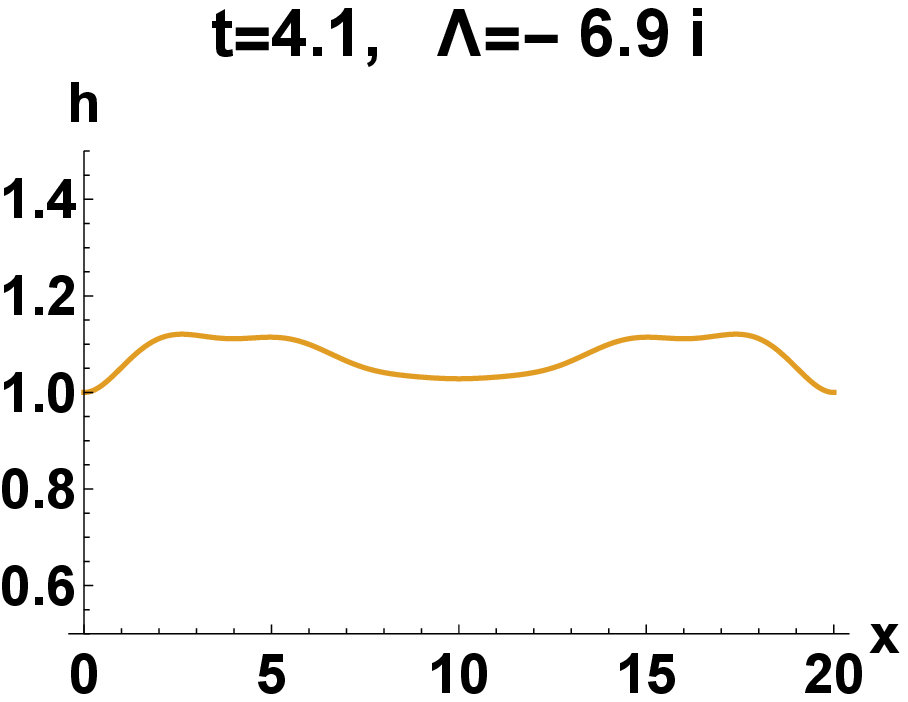}
}
\parbox[]{0.3cm}{(d)\\[2cm](e)\\[2cm](f)}
}
}
\caption{
\label{example} 
Time evolution of the one-dimensional fluid interface $h(x,t)$
for $G=1,\Gamma=1.2, H=0.05$. The initial condition of all simulations is shown in (a) and corresponds to a resting fluid with an elevation in the center. 
(b)-(c) show two snapshots of the time evolution of the interface without external current. In (d)-(e) two snapshots of the onset of unstable behaviour
for $D=-0.2$ according to (\ref{insosc}), and in (f) one snapshot of stable oscillating behaviour $D=0.2$ according to (\ref{osc}) are selected. 
The corresponding wave lengths $\Lambda$ are given above.}
\end{figure}

The unstable oscillatory behaviour leads to the condition for the external current from (\ref{af}) 
for our parameter regime (\ref{para}) and (\ref{par}) 
\be
2 H-2 \sqrt{\Gamma +{G\over k^2}}<D<-2 H.
\ee 
This range can be resolved alternatively also with respect to  the wavelengths
\be
&&\left (2(H-\sqrt{\Gamma})<D<-2H, \forall k\right ) 
\nonumber\\
&&\vee \left ( D<2 (H-\sqrt{\Gamma}), k^2<{4 G\over (2H-D)^2-4 \Gamma}\right ). 
\label{insosc}
\ee

A damped oscillation we obtain analogously for
\be
-2 H<D<2 H +2 \sqrt{\Gamma +{G\over k^2}}.
\ee 
or resolved with respect to the wavelength
\be
&&\left (-2H<D<2(H+\sqrt{\Gamma}), \forall k\right ) 
\nonumber\\
&&\vee \left ( 2 (H+\sqrt{\Gamma})<D, k^2<{4 G\over (2H-D)^2-4 \Gamma}\right ). 
\label{osc}
\ee
This region of damped or increasing oscillations corresponding to $D\gtrless0$ we rewrite from the dispersion (\ref{disp}) as
\be
\delta h,\delta \Phi\sim {\rm e}^{-i \omega(k) t+i \V k \V r}={\rm e}^{-{\alpha(k) t}\pm i t\Omega(k)}
\ee
with the real wave-number-dependent 
damping rate $\alpha(k)$ and frequency $\Omega(k)$ of
\ba
\alpha(k)\!=\!k^2\!\left (\!H\!+\!{D\over 2}\right ),\, \Omega(k)=k\sqrt{G\!+\!\Gamma k^2\!-\!k^2\left (H\!-\!{D\over 2}\right )^2}
\end{align}
respectively.
We see that the viscosity as well as the external current contributes to the damping.

\begin{figure}[h]
\includegraphics[width=6cm]{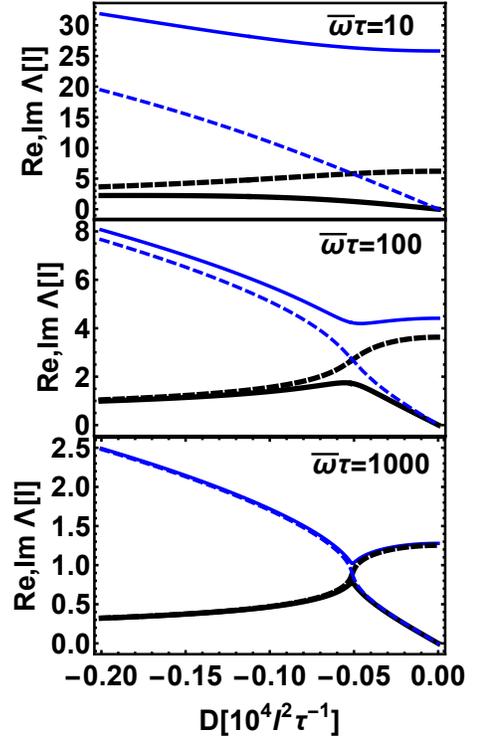}
\caption{\label{kw} The real (solid) and imaginary (dashed) part of the wavelength (\ref{kk}) in dependence on the external current for three different frequencies and the parameters of (\ref{par}).}
\end{figure}

\subsection{Evanescent waves}
In order to see which physical wave is modified here by the various parameters, we can solve (\ref{disp}) alternatively for the wave vector now dependent on the real frequency
\be
\delta h,\delta \Phi\sim {\rm e}^{-i \omega t+i \V k(\omega) \V r}
\ee
to obtain
\ba
k^2\!&=\!{i \omega (D\!+\!2 H)\!-\!G\!\pm\! \sqrt{[i\omega (D\!+\!2 H)\!-\!G]^2\!+\!(8 H D\!+\!4\Gamma) \omega^2}\over 4 HD\!+\!2\Gamma}
\nonumber\\
&\approx {i \omega D-G\pm \sqrt{(i\omega D-G)^2+4\Gamma \omega^2}\over 2\Gamma}
\nonumber\\
&=\left \{ \begin{array}{cc}
-{\omega^2\over G}+{-G+i D\omega\over \Gamma},\,{\omega^2\over G}&+o(\omega^3)
\cr
&
\cr
-{G\over 2\Gamma}\left (1\!\pm\! i{D\over\sqrt{4 \Gamma\!-\!D^2}}\right )\!\pm\! {\omega\over 2\Gamma}(iD\pm \sqrt{4 \Gamma-D^2})&\!+\!o(\omega^{-1})
\end{array}\right . 
\label{kk}
\end{align}
where the viscosity damping $H$ is omitted as being small according to (\ref{par}) in the second line. There are two regimes according to the size of frequency. We see from the case of small frequencies without damping and external perturbation that we have just gravitational waves with the phase velocity $\sqrt{g h_0}$. This will become modified strongly by viscosity and surface tension coupling. 

The opposite limit of large viscosity $H$ to surface tension $\Gamma$ limit reads from (\ref{kk})

\ba
k^2\!&\approx \left \{ 
\begin{array}{cc}
-{G\over 2 H J}+{i (J+2 H)\omega\over 2 H J}&+o(\omega^2)
\cr
&
\cr
{G\over J(J-2H)}-i{\omega\over J},{G\over 2 H(2H-J)}-i{\omega\over 2 H} &\!+\!o(\omega^{-1})
\end{array}\right .. 
\label{kk1}
\end{align}
One sees that for small frequencies evanescent waves appear and for large frequencies that two modes appear where the surface tension and external current acts alternatively. These are capillary waves as one sees from the dispersion relation (\ref{af})  
for small viscosity and neglecting the external current
\be
\omega^2=k^2(G+\Gamma k^2)
\ee
which can be compared with the standard expression for gravity–capillary waves $\omega^2=(g k+{\gamma\over n} k^3) tanh(k h)\approx h k^2(g+{\gamma\over n}k^2)$ of a liquid \cite{Ph77}. The laser-induced capillary wave have been treated in \cite{AH87}. 

Let us now discuss the full expression (\ref{kk}) including viscosity and surface tension. In figure~\ref{example} some numerical snapshots for one dimension are given of the time evolution of (\ref{vel}) and (\ref{vel2}) together with their resulting wavelength (\ref{kk}). We see for the case without external current in (a)-(c) how an initial disturbance is decaying into an evanescent wave with the corresponding wave length of $6.9$. With the same initial disturbance we consider the influence of external current for the unstable oscillating $D=-0.2$ and stable oscillating $D=0.2$ behaviour according to (\ref{insosc}) and (\ref{osc}) respectively. Two snapshots (d) and (e) illustrate the onset of unstable oscillations and (f) a time instant of the stable case. 

In figure~\ref{kw} we plot the dependence of the real and imaginary parts of the wavelength $k=2\pi/\Lambda$ in dependence on the external current and frequency of the resulting wave. The real part is even and the imaginary part is odd as function of frequency. We see that due to the external current the real part of the wavelength is first reduced and than increases linearly such that we can scale (\ref{lam0}) linearly with the dimensionless external current. 

\begin{figure}[h]
\includegraphics[width=4.2cm]{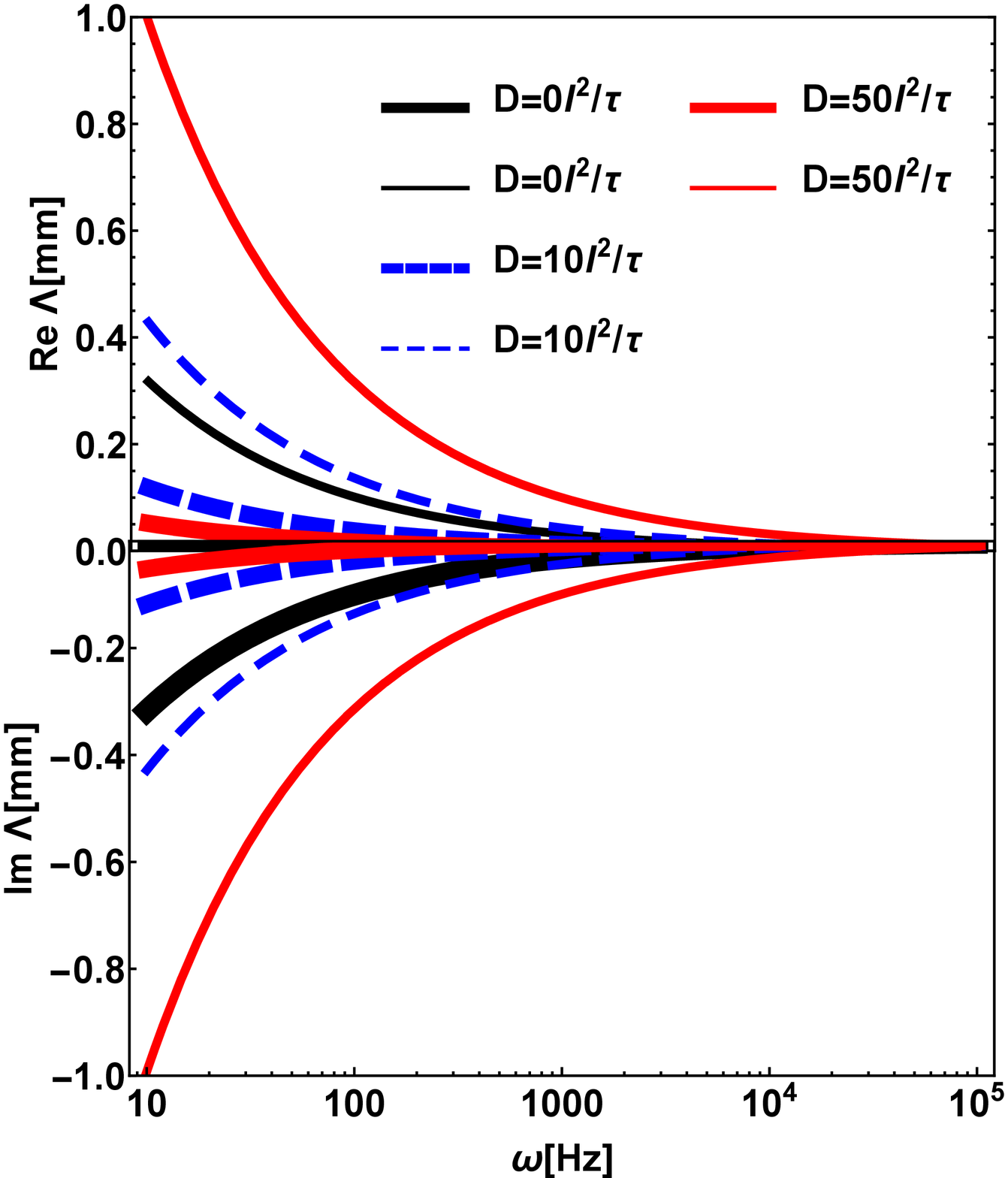}
\includegraphics[width=4.2cm]{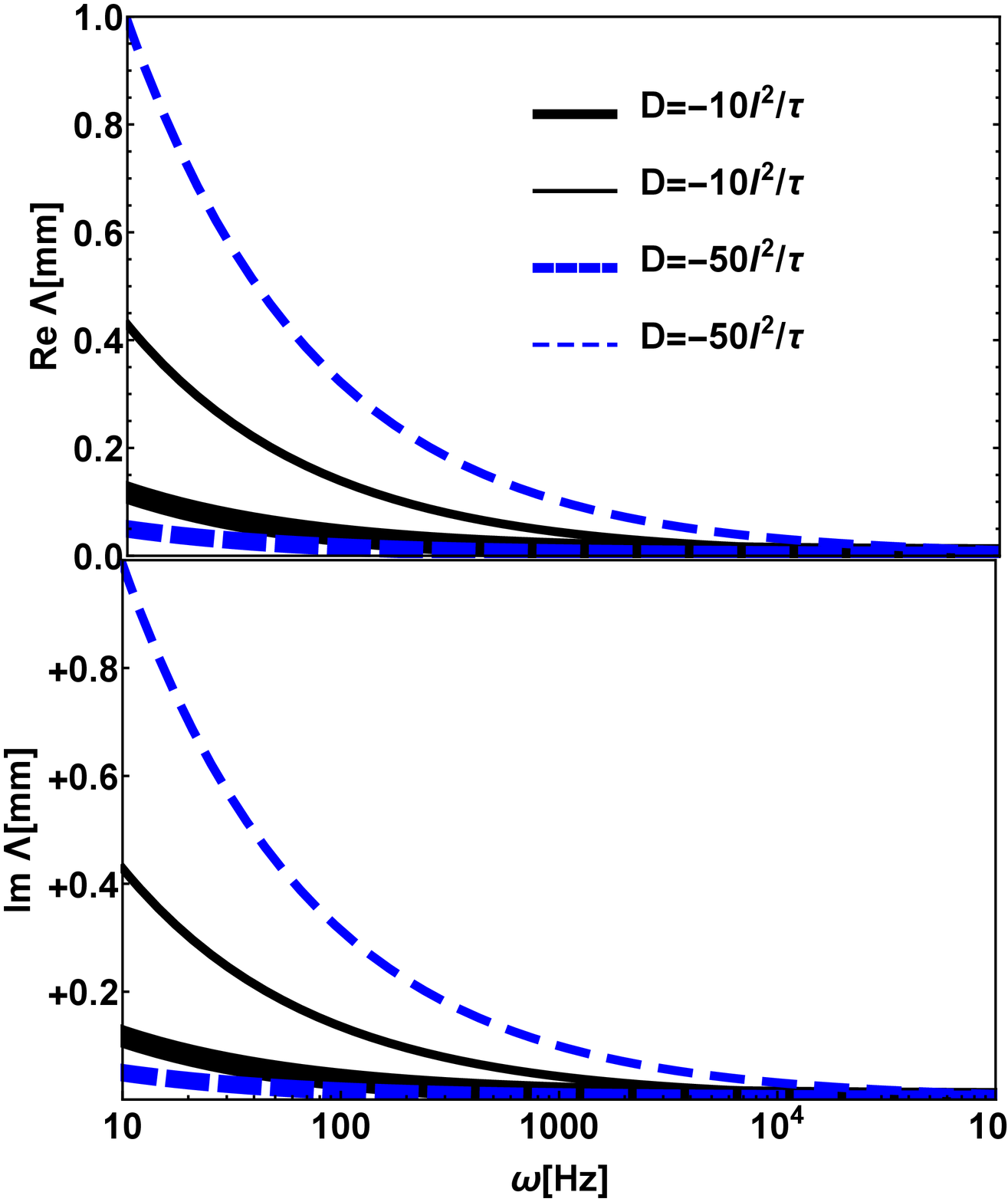}
\caption{\label{lw} Left: stable (damped) case $D\ge0$, Right: unstable case $D<0$ of the two wavelengths (\ref{kk}) in dependence on the frequency for liquid $Au$ with parameters of (\ref{typ}). The corresponding real parts (above) and corresponding imaginary parts (below).}
\end{figure}

\begin{figure}[h]
\includegraphics[width=4.2cm]{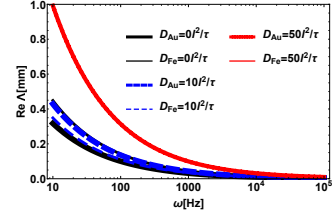}
\caption{\label{lw_Au_Fe} Comparison of $Au$ and $Fe$ for the corresponding larger wavelength of figure~\ref{lw}.}
\end{figure}

For the parameter of $Au$ (\ref{typ}) we plot the real and imaginary parts of the wavelength in figure~\ref{lw}. There are two modes. Without external current the smaller mode with respect to the real part is more damped than the larger mode which turns into the opposite for larger external currents. Here the larger mode is much more damped. The unstable case $D<0$ is accompanied by positive imaginary parts corresponding to growth rate as in figure~\ref{kw} . The range of ripple formation can reach $mm$ dependent on the frequency and external current which is in the observed range. The comparison between $Au$ and $Fe$ parameters in figure~ \ref{lw_Au_Fe} reveals that the modes are different only for small external beams. For larger beams the difference becomes negligible. 

\subsection{External frequency dependence\label{Floquet}}

\begin{figure}[h]
\parbox[]{8.8cm}{
\includegraphics[width=4.5cm]{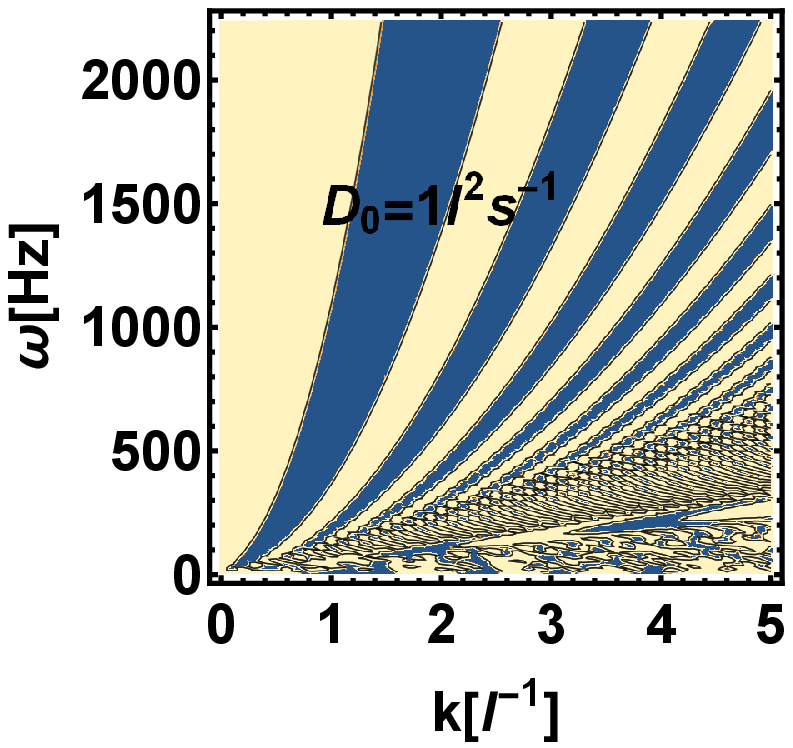}
\includegraphics[width=3.35cm]{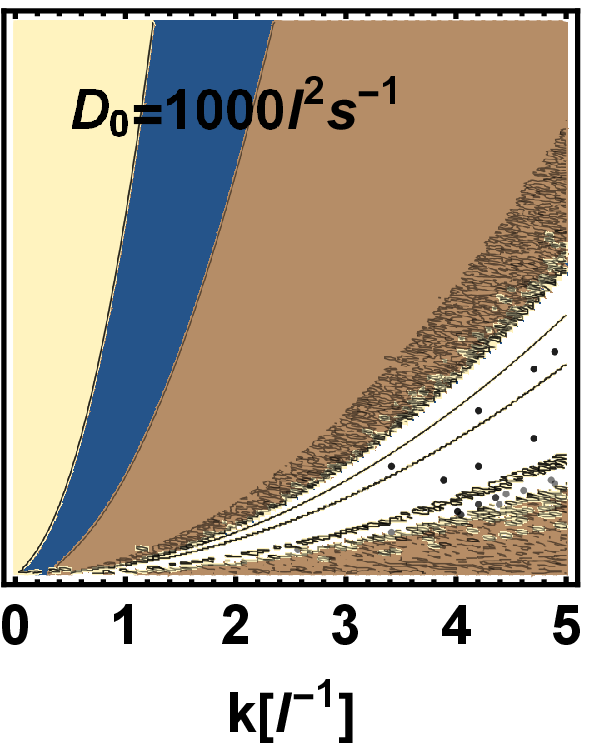}
\includegraphics[width=0.5cm]{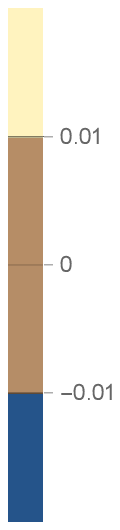}
}
\caption{\label{floquet} The contours of Floquet multipliers showing the  alternating border between stable (blue) and unstable (yellow) behaviour for an external frequency in dependence on the wavevector and two different external beam amplitudes. The white area indicates the range which is marginal stable with $\nu_i=0$. The uppermost left parabola limits the range of oscillatory instabilities. The parameters for $Au$ according to (\ref{par}) are chosen independent on characteristic time $\tau$. }
\end{figure}

So far we have considered constant external currents which means the frequency of the linear response is the own frequency of the system created by the interplay between gravitation, surface tension and viscosity. As soon as the external current imposes a certain frequency the situation becomes more complicated since now the periodic time dependence of $D(t)=D(t+T)=D_0\sin(2\pi t/T)$ will create perturbation which stability can be analyzed with the help of Floquet theory \cite{FI99,KK14}. We numerically solve the equation system (\ref{disp}) in time domain to determine the fundamental matrix 
\be
X(t)=\begin{pmatrix}
\delta h_1(t)&\delta h_2(t)\cr
\delta \Psi_1(t) &\delta \Psi_2(t)
\end{pmatrix}
\ee
where any linear combination of the fundamental solutions $(\delta h,\delta \Phi)_{1,2}$ solves the equation system. Therefore we can arrange for $X(0)=I$ which means we solve (\ref{disp}) with this initial conditions. Then the eigenvalue of the monodromy matrix or Poincar\'e mapping 
\be
C=X^{-1}(0)X(T) 
\ee
yields the Floquet multipliers $\nu_i$ which determine the stable $\nu_i<0$ and unstable $\nu_i>0$ behaviour. In figure~\ref{floquet} we plot the region of instability for a certain amplitude $D_0$ in dependence on the wavevector of $\delta h$ and $\delta \psi$. One sees that the borders between stable and unstable behaviour is here nearly a quadratic curve alternating with increasing wavevector. This quadratic behaviour follows the linear response result. Larger external beams increase these regions. The uppermost left parabola limits the range of unstable oscillations while all other regions show a single resonance peak which decreases or increases with time according to the sign of Flouquet parameters. 

The external frequency we use in figure~\ref{floquet} is not the one which the system will develop as combination of both external and internal one in the upper left unstable parabola. Instead of analyzing this resulting frequency some estimates from linear response should be sufficient to discuss possible scenarios.  For small resulting frequencies assuming that it is given by the repetition rate of $200kHz$
with the values (\ref{par}) we 
obtain  from (\ref{kk}) besides a normal wave also evanescent waves which are induced by the external current. 
The wave length of the normal wave without viscosity, surface tension and external current reads then in dependence on the frequency 
\be
\Lambda_0&=&{2\pi \sqrt{G}\over \omega} l={2\pi \sqrt{gh_0}\over \bar \omega}\nonumber\\
&\approx& {2\pi 10^2 {\tau \over s} \over \omega [Hz] {\tau \over s}} 10^{-3}m={10^{-1} m\over {\omega[Hz]\over 2\pi}}
\label{lam0}
\ee
where we reintroduced the dimension-full frequency $\bar \omega=\omega/ \tau$ and use the repetition rate of the laser of $200$ kHz for $\omega/2\pi$.
We can estimate this wave length for $Au$ according to the parameters (\ref{para}) and table~\ref{TableParam} as being
$\Lambda_0\approx 0.5 \mu m$. This free result will become strongly modified now by the damping and the external current as we discussed in figure~\ref{lamD}. 

The other regime of large frequencies we might apply if the frequency would be thought of as given by the laser light. We can estimate with (\ref{typ}) assuming a wavelength of the initial laser of $1000$nm
\be
\omega=\bar \omega \tau=c \tau k_0\approx 10^{13}
\ee
which would lead to an unrealistic wavelength of ripples of
\be
k\!=\!\!\sqrt{\omega\over \sqrt{\Gamma}}\to \Lambda\!=\!{2\pi l\over k}\!=\!{2\pi (6.5)^{1/4}10^{-3}\over  4 \times 10^{6} }m\approx 3nm.
\ee 
Hence, the observed ripple formation cannot be due to direct electromagnetic coupling of the laser light to the surface as assumed in the literature for laser impact on semiconductors. A stationary interference between scattered light from the surface and the cavity radiation as proposed in \cite{EHW73} seems to be unlikely since the first impact of laser melts the smooth surface and a followed radiation interference on the surface, which topology changes pulse-to-pulse \cite{GG14}.
Therefore we propose that the ripple formation is due to the internal frequency as interplay of gravitation, viscosity and surface tensions triggered by the external frequency which could be the repetition or sweep rate of the laser which means of mechanical origin rather than electromagnetical origin.

\section{Weak nonlinear stability and structure analysis}

\subsection{Possible stable structures}

\begin{figure}[h]
\includegraphics[width=7cm]{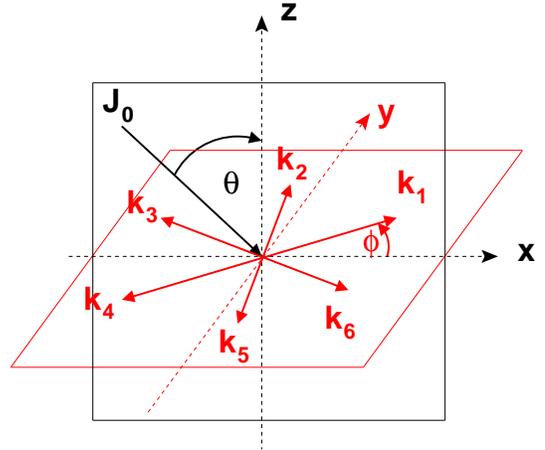}
\caption{\label{star} Sketch of incident plane $x,z$ (black) given by the incoming beam $J_0$ under incident angle $\Theta$ together with the surface $x,y$ (red) and the geometry of amplitude analysis used for  hexagonal structures where $\phi$ is the angle to the beam-x axes at the surface.}
\end{figure}

We can decide for which parameters quadratic, hexagonal or stripe structures
will appear. Therefore one represents the structure by four or six wave vectors, respectively, which are pairwise oppositely directed. The amplitudes belonging to the
pairwise wave vectors are complex conjugated to each other in order to render
the ansatz
\be
\Phi_{00}({\bf r},t)=\sum\limits_i A_i(t) {\rm e}^{i{\bf k_i x}};\quad k_i^2=1
\label{ansatz}
\ee
real illustrated in figure~\ref{star} for hexagons and specially for stripes with $A_1=A_2=0,A_3\ne 0$. Analogously we expand $h(\V r,t)$ into $B_i$ and $f(\V r,t)$ into $F_i$ amplitudes
\be
h({\bf r},t)&=&h_0+\sum\limits_i B_i(t) {\rm e}^{i{\bf k_i x}}
\nonumber\\
f({\bf r},t)&=&f_0+\sum\limits_i F_i(t) {\rm e}^{i{\bf k_i x}}
\label{ansatz1}
\ee
where with our scaling $h_0=1$.

Introducing the ansatz (\ref{ansatz}) into the nonlinear equations (\ref{sys}), multiplying
with ${\rm e}^{-i{\bf k_l x}}$, and integrating over ${\bf x}$ leads to coupled
equations for the amplitudes. One sees that quadratic terms cannot yield quadratic
structures since it leads to ${\bf k_i+k_j-k_l}=0$ as condition which
cannot be completed by two pairwise oppositely directed wavevectors. The
hexagonal structure can be achieved since three pairwise oppositely directed
wavevectors form a hexagon and one has $k_i+k_j-k_l=0$ as the only
possibility to combine three vectors. 

The resulting amplitude system reads 
\ba
\dot A_1&=- (G+\Gamma) B_1-2 H A_1-\frac 1 2 A_2 A_3^*\nonumber\\
\dot B_1\!-\!\dot F_1&=-c_1 B_1+\frac 1 2 (B_2\!-\!F_2) A_3^* \!+\!\frac 1 2 (B_3^*-F_3^*) A_2
\label{sys1}
\end{align}
with cyclic indices $1,2,3$. We have introduced the abbreviation 
\ba
c_1&=D_x k_{1x}^2+D_y k_{1y}^2=D_x \cos^2{\phi}+D_y \sin^2{\phi}
\nonumber\\
c_2&=D_x k_{2x}^2\!+\!D_y k_{2y}^2\!=\!D_x \cos^2{(\phi\!+\!{\pi\over 3})}\!+\!D_y \sin^2{(\phi\!+\!{\pi\over 3})}
\nonumber\\
c_3&=D_x k_{3x}^2\!+\!D_y k_{3y}^2\!=\!D_x \cos^2{(\phi\!+\!{2\pi\over 3})}\!+\!D_y \sin^2{(\phi\!+\!{2\pi\over 3})}
\label{C1}
\end{align}
with the incident-angle dependent coefficients given either by the collision model (\ref{comp}) or
by the surface impingement model (\ref{comp1}). These coefficients could be time dependent if the external current is time dependent. 

\begin{figure}[h]
\includegraphics[width=8cm]{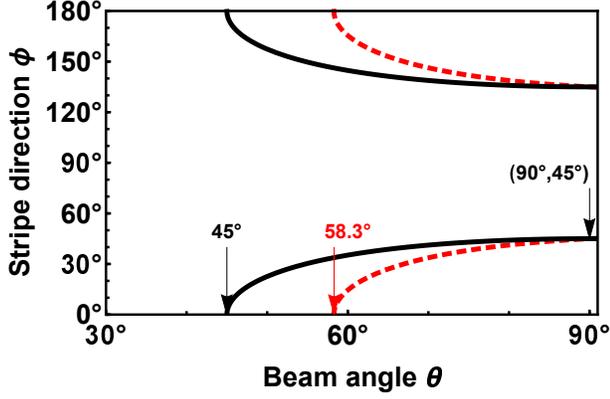}
\caption{\label{stripes} The angle of stripe orientation $\phi$ in the surface as in figure \ref{star} as function of the incident angle $\theta$ of incoming beam for the collision model (\ref{comp}) (dashed) and the surface impingement model (\ref{comp1}) (solid). }
\end{figure}

In case one finds a static solution of (\ref{sys1}) one wants to see the stability around this stationary solution
\ba
&A_1=\bar A_1+\epsilon_1 {\rm e}^{\lambda t}, A_2=\bar A_2+\epsilon_2 {\rm e}^{\lambda t}, A_3=\bar A_3+\epsilon_3 {\rm e}^{\lambda t},
\nonumber\\
& B_1=\bar B_1+\epsilon_4 {\rm e}^{\lambda t}, B_2=\bar B_2+\epsilon_5 {\rm e}^{\lambda t}, B_3=\bar B_3+\epsilon_6 {\rm e}^{\lambda t}.
\label{stab}
\end{align}
We will analyze the internal possible structure of the system and assume no external time dependence $F(t)=const$ and $D(t)=const$.  
The possible growth rates $\lambda$ are then the solutions of the eigenvalue problem to the matrix
\ba
\left(\!\!
\begin{array}{cccccc}
 -4 H & -A_3 & -A_2 & -2 (\Gamma \!+\!G) & 0 & 0 \\
 -A_3 & -4 H & -A_1 & 0 & -2 (\Gamma \!+\!G) & 0 \\
 -A_2 & -A_1 & -4 H & 0 & 0 & -2 (\Gamma \!+\!G) \\
 0 & B_3 & B_2 & -2 c_1 & A_3 & A_2 \\
 B_3 & 0 & B_1 & A_3 & -2 c_2 & A_1 \\
 B_2 & B_1 & 0 & A_2 & A_1 & -2 c_3 \\
\end{array}\!\!
\right).
\label{det}
\end{align}
Stable structures demand that all growth rates $\lambda$ are negative.

\subsection{Stripe formation}

We analyse the structure of solutions for the special case that the ground is shaped in 1-direction and $F_2=F_3=0$. We search for the stripe solution $A_2=A_3=0$. Eq.s (\ref{sys1}) provide the conditions
\be
c_1=0,\, B_1=-{2 H\over G+\Gamma} A_1,\, B_2=B_3=0.
\ee
The first one, $c_1=0$ leads with (\ref{C1}) to a relation between the incident angle $\theta$ of the incoming beam and the orientation angle $\phi$ of stripes
\be
\tan^2\phi=-{D_x\over D_y}=\left \{
\begin{array}{cc}
-\cos{(2\theta)}-\sin{\theta}\cos{\theta}&(\ref{comp})
\cr
-\cos{2\theta}&(\ref{comp1})
\end{array}
\right ..
\ee
The other two constants take the values
\be
c_{2/3}=D_y \left [\frac 3 4 (1-\tan^2\phi)\pm {\sqrt{3}\over 2} \tan\phi \right ].
\ee

The results are given for both models, collisional model (\ref{comp}) and surface impingement model (\ref{comp1}) in figure~\ref{stripes}. One sees that in the collision model the beam incident angle has to be larger than $58.3^o$ to form a stripe structure. The surface impingement model leads to a minimal angle of $45^o$. The maximal angle between the incoming plane $x$-direction and the stripe orientation can reach $45^o$ at a perpendicular beam for both models.

\begin{figure}[]
\includegraphics[width=4cm]{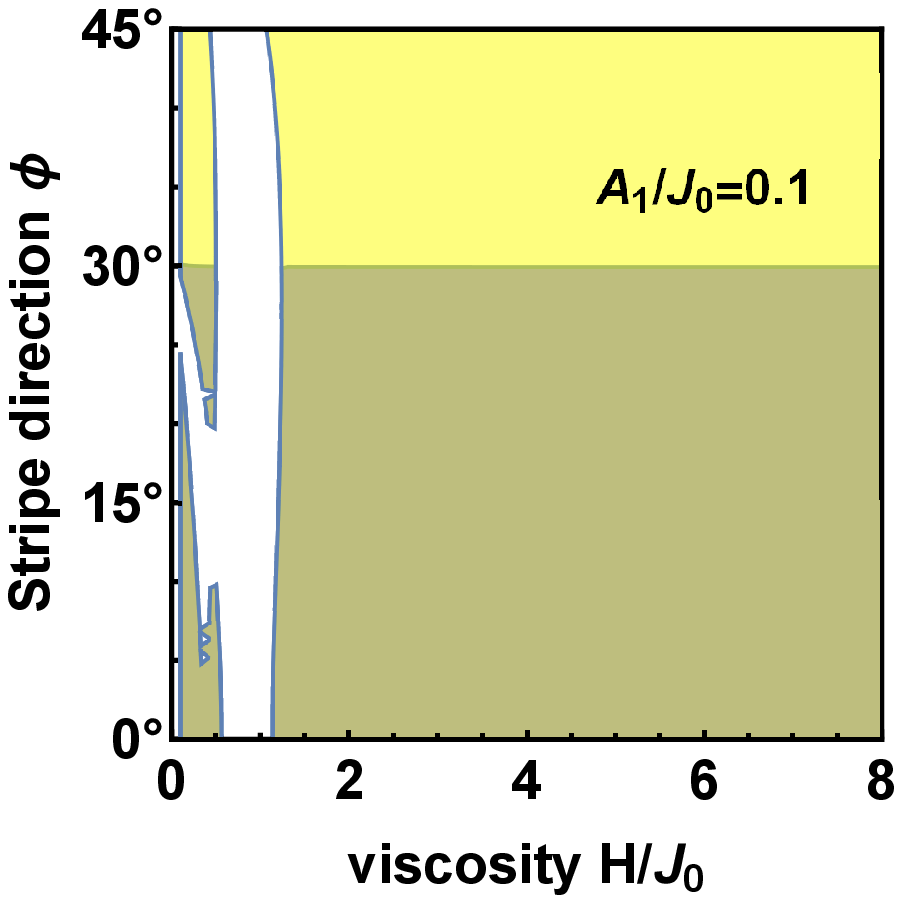}
\includegraphics[width=4cm]{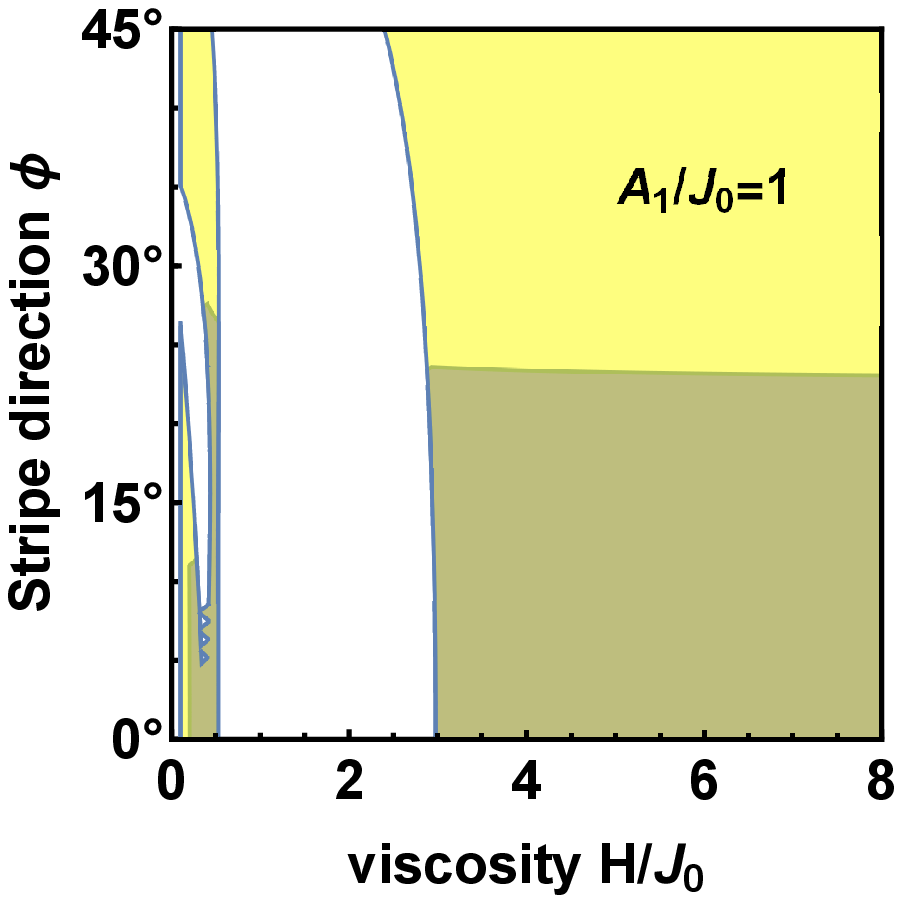}
\includegraphics[width=4cm]{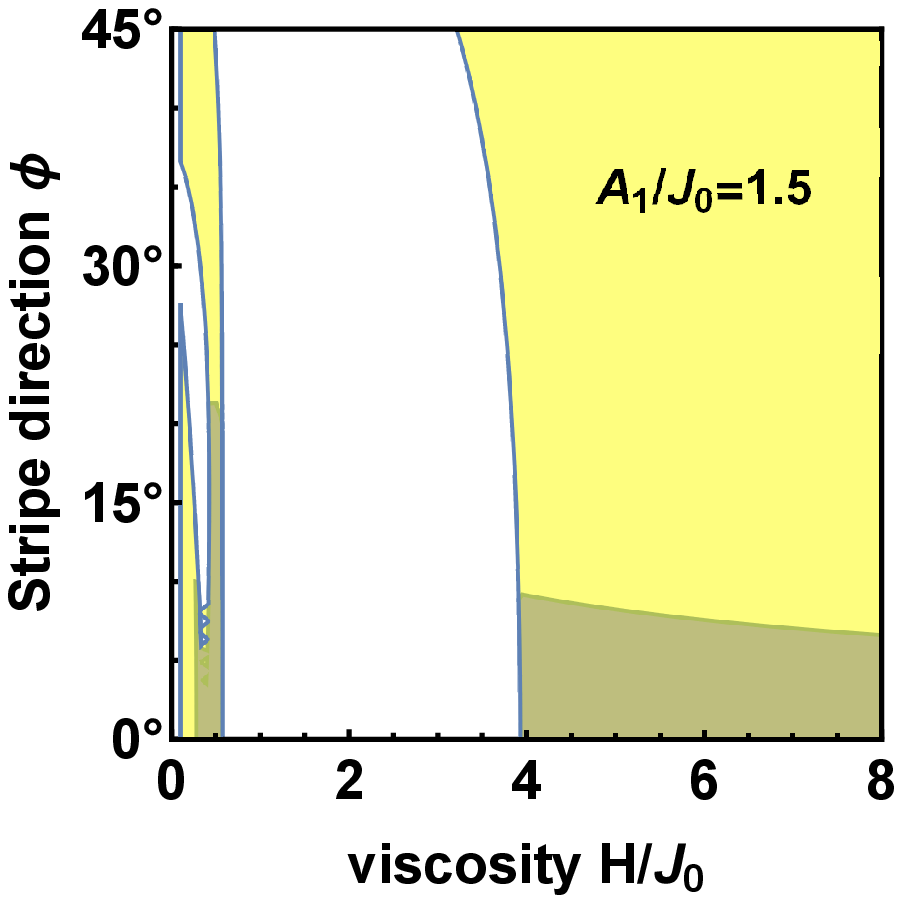}
\includegraphics[width=4cm]{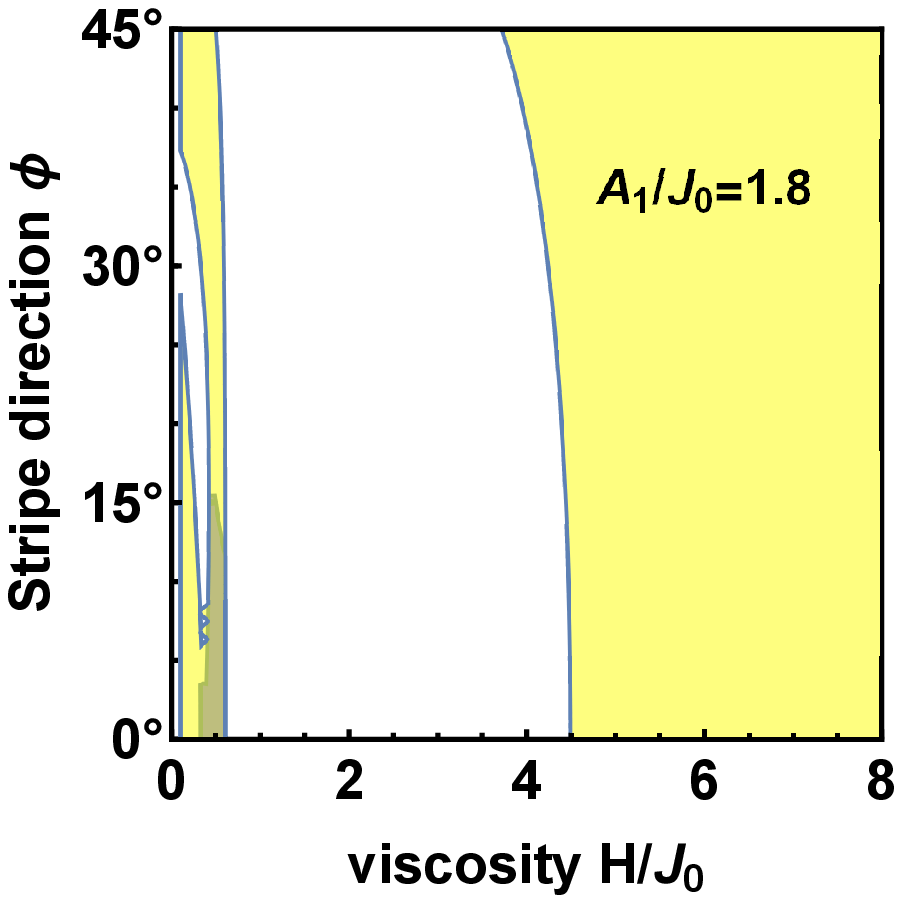}
\caption{\label{stripes_range} The range of possible stable stripes as function of the viscosity and the angle of stripe orientation with respect to the plane of incoming beam for four different amplitudes of the collisional model. 
The area with $Re\lambda<0$ is indicated by blue and the white area represents oscillating behaviour. We choose $G+\Gamma=0.1$ which parameter almost does not influence the result. The amplitudes and viscosity parameter are scaled in terms of external current.}
\end{figure}


\begin{figure}[h]
\parbox[]{8.7cm}{
\parbox[]{4.cm}{
\includegraphics[width=4cm]{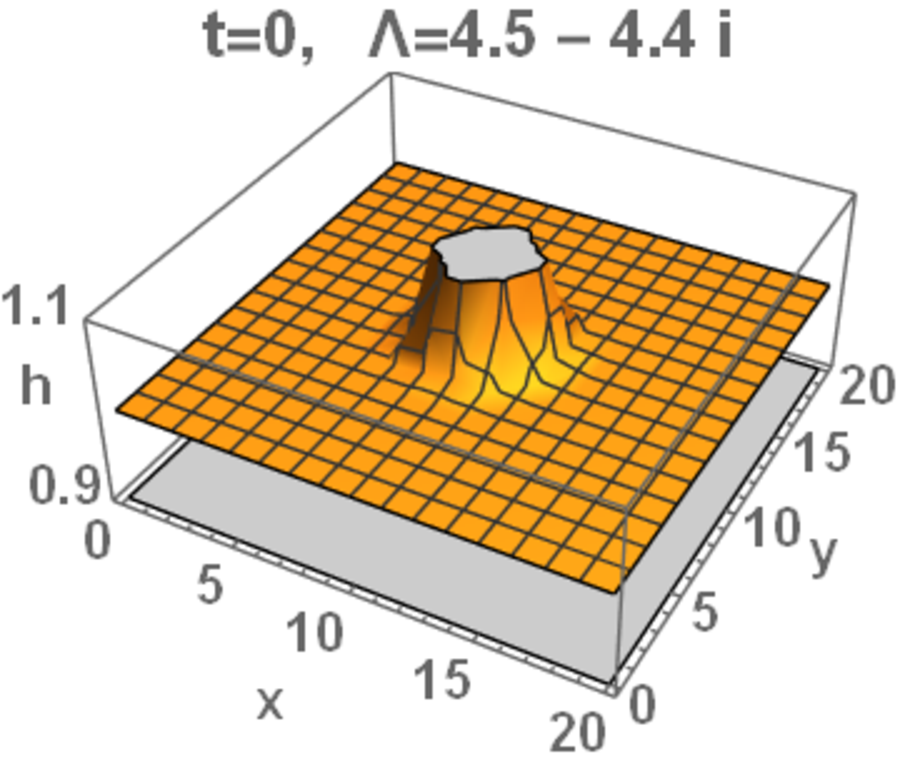}
\includegraphics[width=4cm]{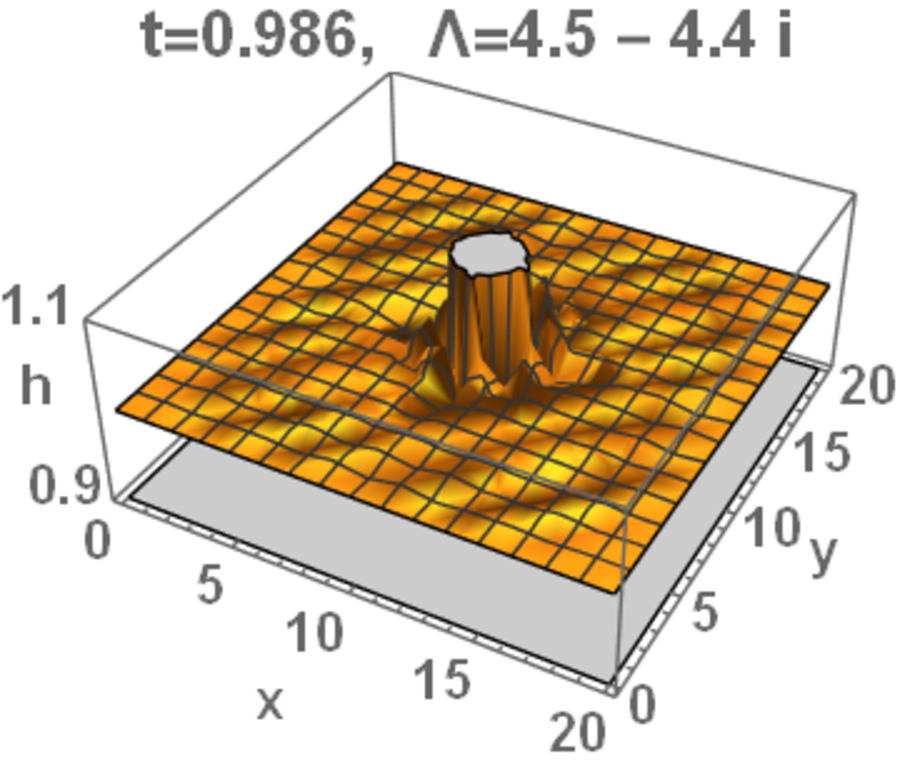}
\includegraphics[width=4cm]{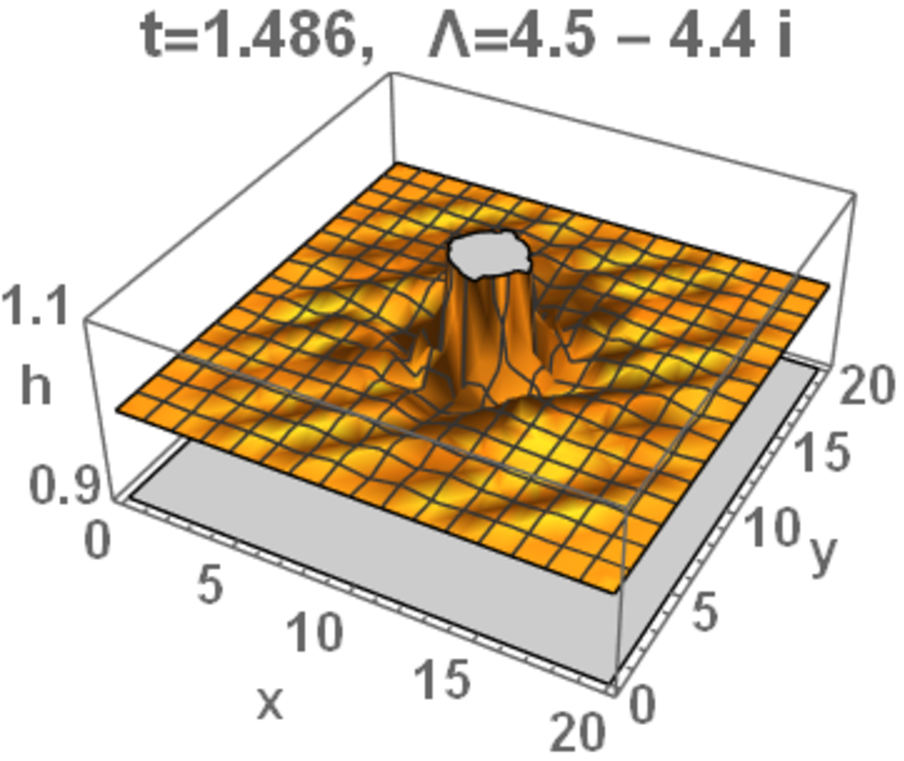}
}
\hspace{0.5cm}
\parbox[]{4.cm}{
\includegraphics[width=4cm]{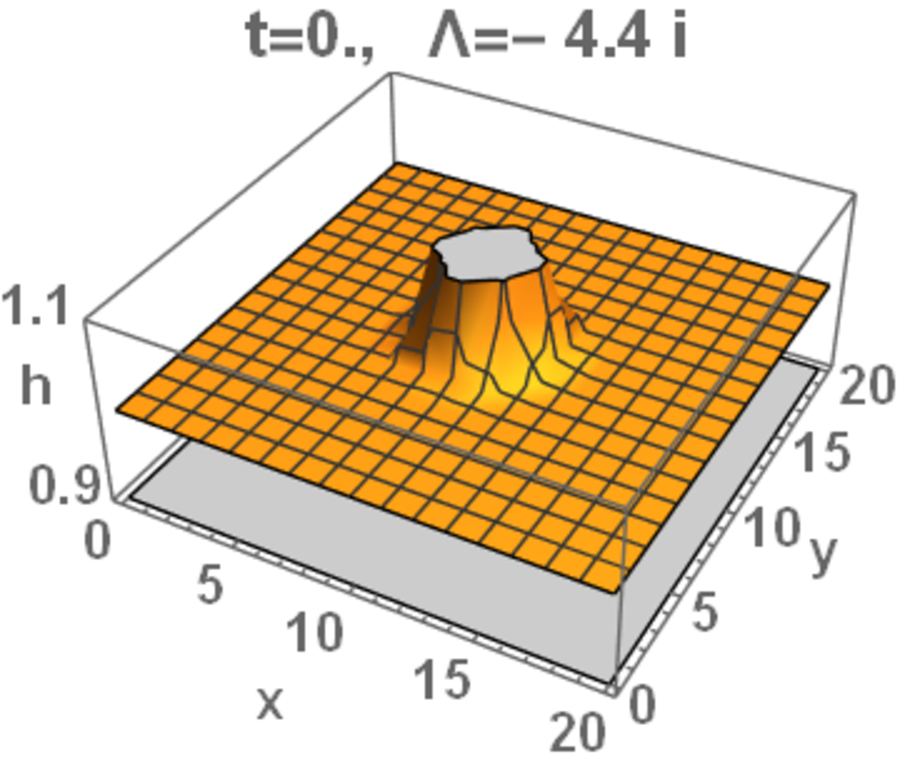}
\includegraphics[width=4cm]{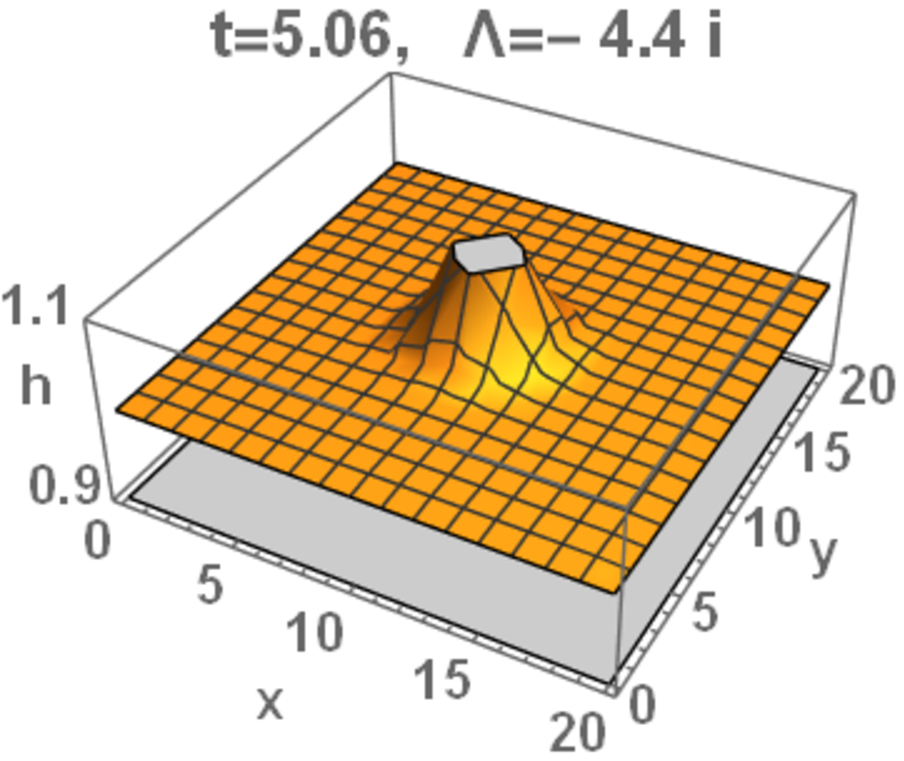}
\includegraphics[width=4cm]{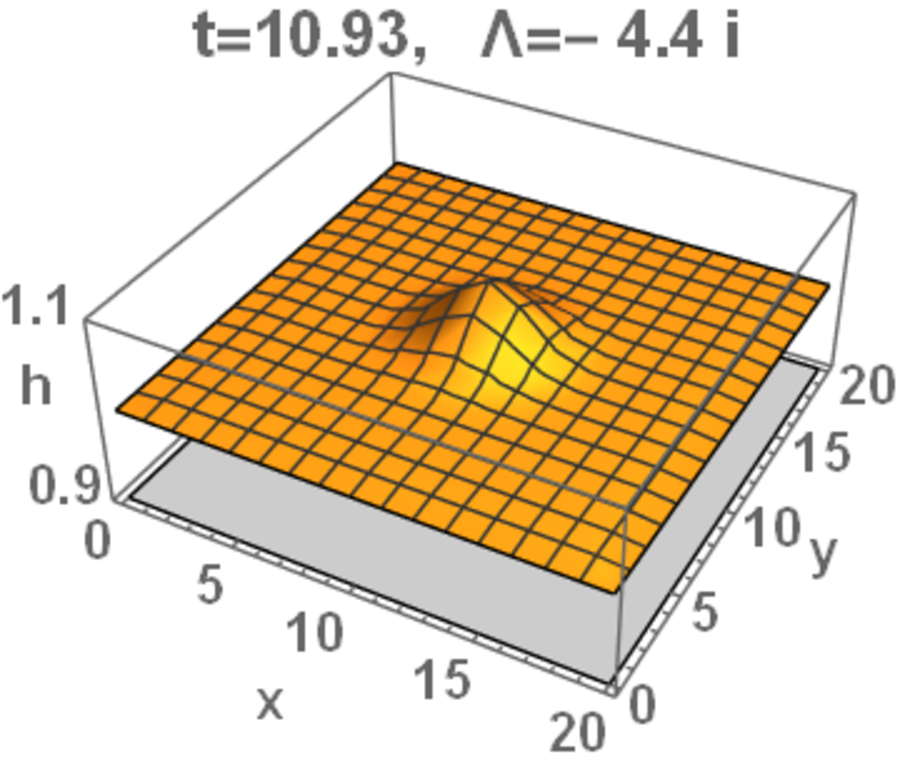}
}
}
\caption{\label{stripes1} Three time steps of the two-dimensional set (\ref{sys}) for $G=1$, $\Gamma=0.5$, $H=5$ with an initially elevated surface. Left side: with external current $D_x=D_y=-2\sin{\omega t}$ and $\omega=10$ corresponding to the unstable ripple formation of the upper left figure in figures~\ref{stripes_range} 
and $\phi=45^o$, right side: without external current. The wavelengths of linear response are given above for comparison.}
\end{figure}


Now we discuss the stability of the stripes and solve the eigenvalues of (\ref{det}). For stripes and without external current we obtain the six solutions
\be
\lambda={0,-2 H, -H\pm\sqrt{A_1^2\pm 8 A_1 H+4 H^2}}
\ee
and demanding $\lambda\le 0$ is only possible for $A_1=0$ which means no structure at all. Therefore we conclude that without external current no stable stripe structure can appear.

This changes if we add the external current. We find stable structures of $Re[\lambda]<0$. Distinguishing between oscillating and stationary structures it is illustrated in figure~\ref{stripes_range} that they become dependent on the amplitude. We obtain stable stripe structures only for amplitudes $A_1<0.4 J_0$ for the impingement model and $A_1<1.8 J_0$ for the collisional model which underlines the importance of external current. The maximal reachable angle for stable stripes is found to be $\approx 30^o$ which restricts the range of available values given by $c_1=0$ and figure~\ref{stripes} further. The dependence on the viscosity is rather weak except that for smaller viscosities damped oscillating structures appear. This range of oscillating behaviour increase up to higher viscosities for larger amplitudes. The dependence of the growth rate of $\Gamma+G$ was not observable within a range of four orders of magnitude. The difference between the collisional model and impingement model is that the latter one restricts the amplitudes to somewhat smaller values.

Three time steps of (\ref{sys}) are seen in figure~\ref{stripes1} for the unstable region with parameters corresponding to the upper left situation of figure~\ref{stripes_range} 
and $\phi=45^o$. The initial disturbance was a Gau\ss{} profile with $A_1/J_0=0.1$. The visible formation of ripples is followed by an exponential growth. The time evolution without external current leads just to a damped decay of the initial disturbance as seen on the right side of figure~\ref{stripes1}.

\begin{widetext}

\begin{figure}[h]
\parbox[]{13.5cm}{
\parbox[]{4.cm}{
\includegraphics[width=4cm]{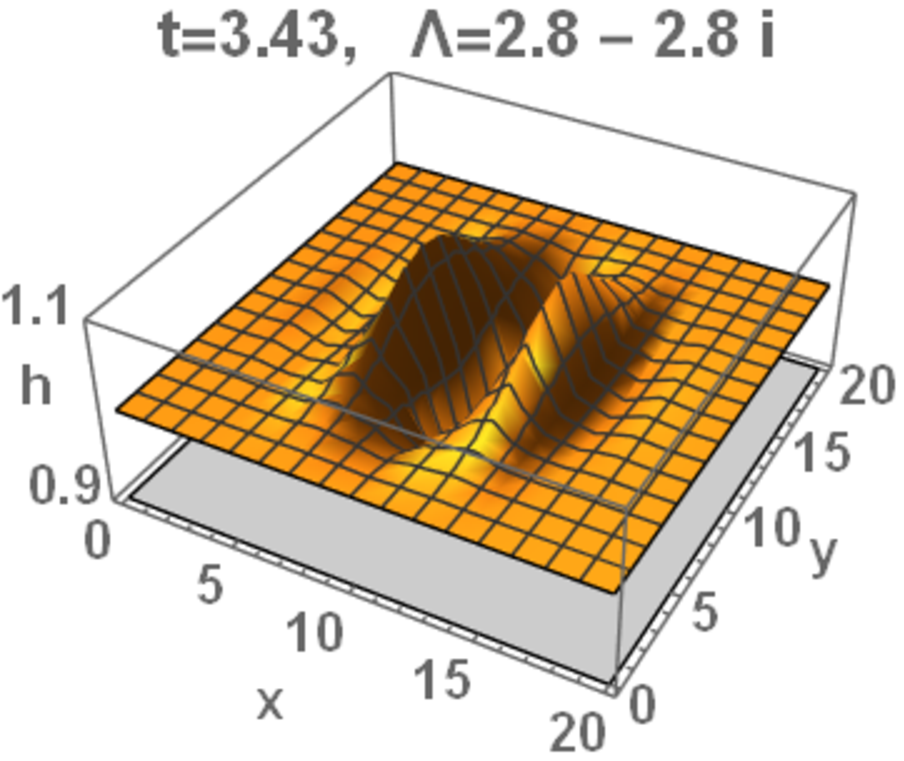}
\includegraphics[width=4cm]{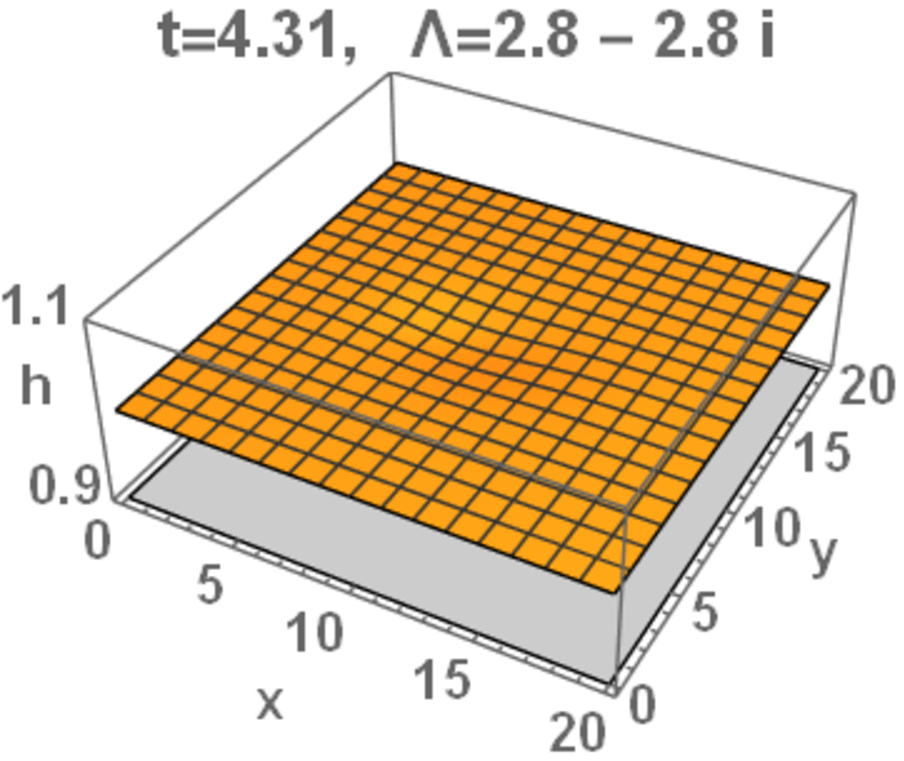}
\includegraphics[width=4cm]{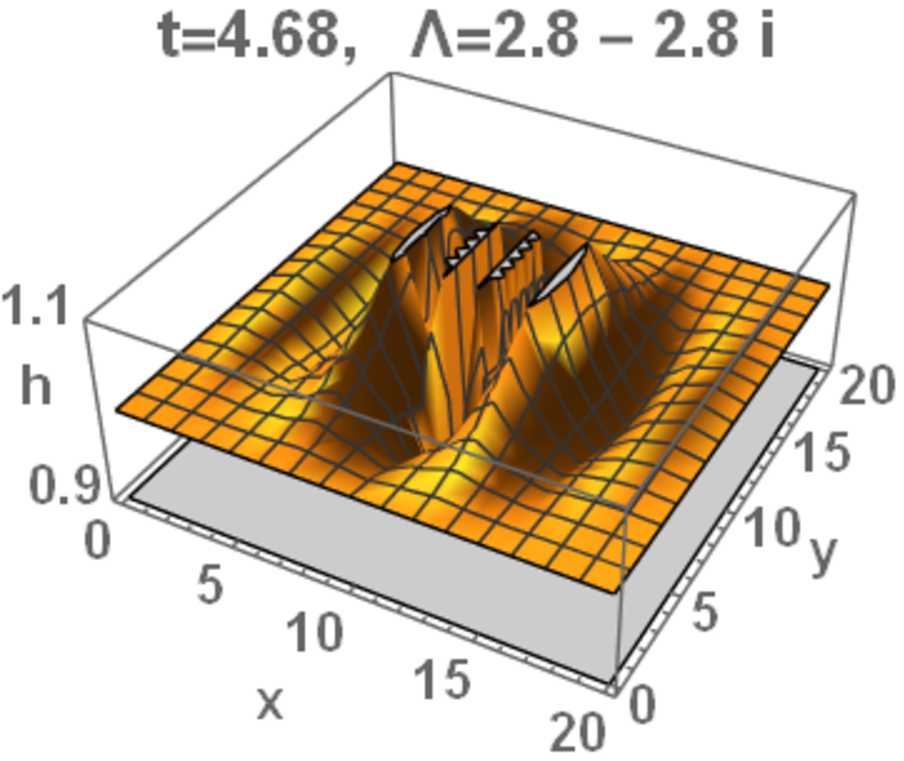}
\includegraphics[width=4cm]{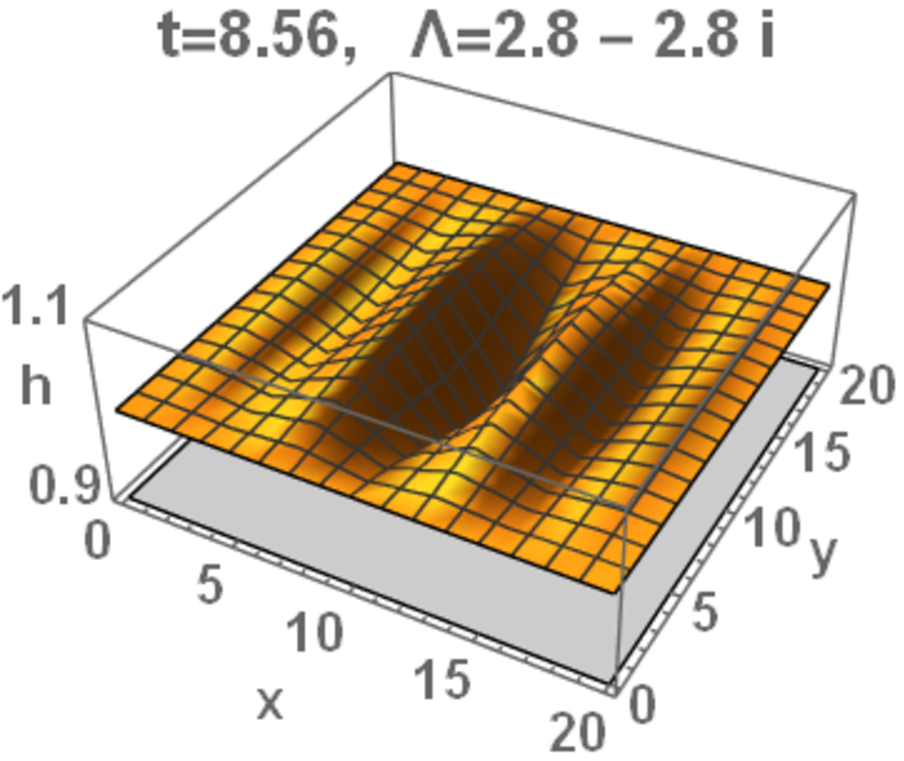}
}
\hspace{0.5cm}
\parbox[]{4.cm}{
\includegraphics[width=4cm]{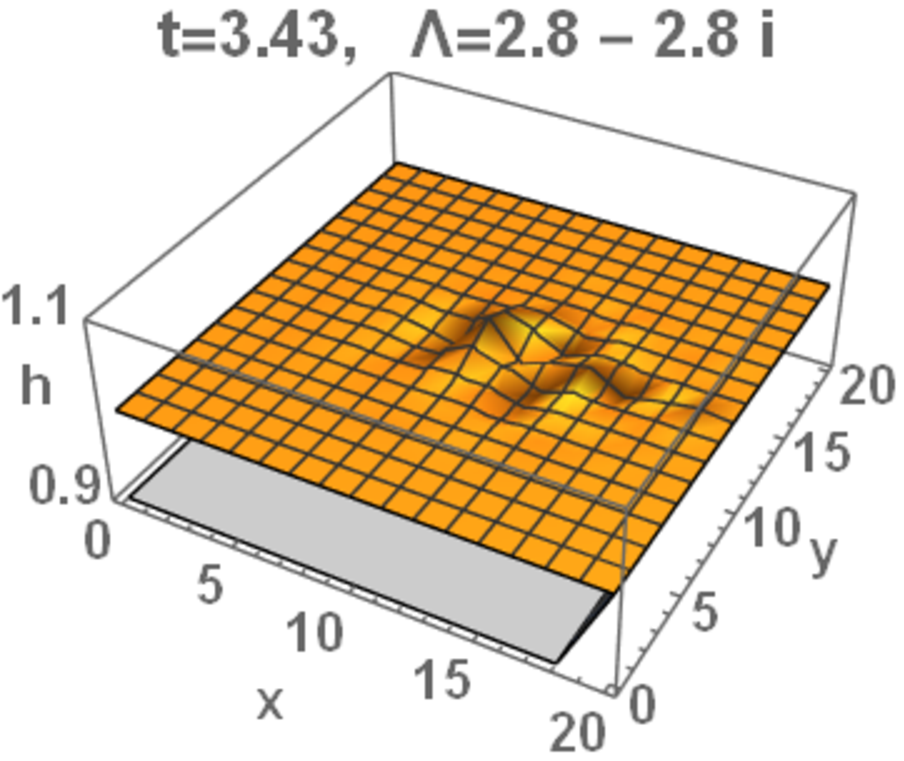}
\includegraphics[width=4cm]{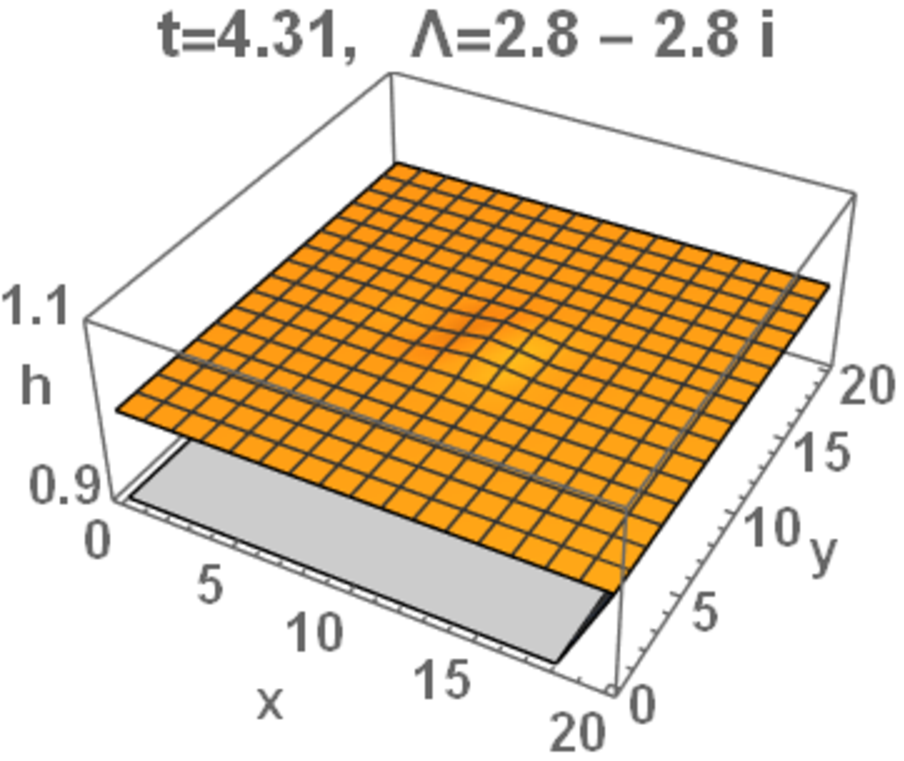}
\includegraphics[width=4cm]{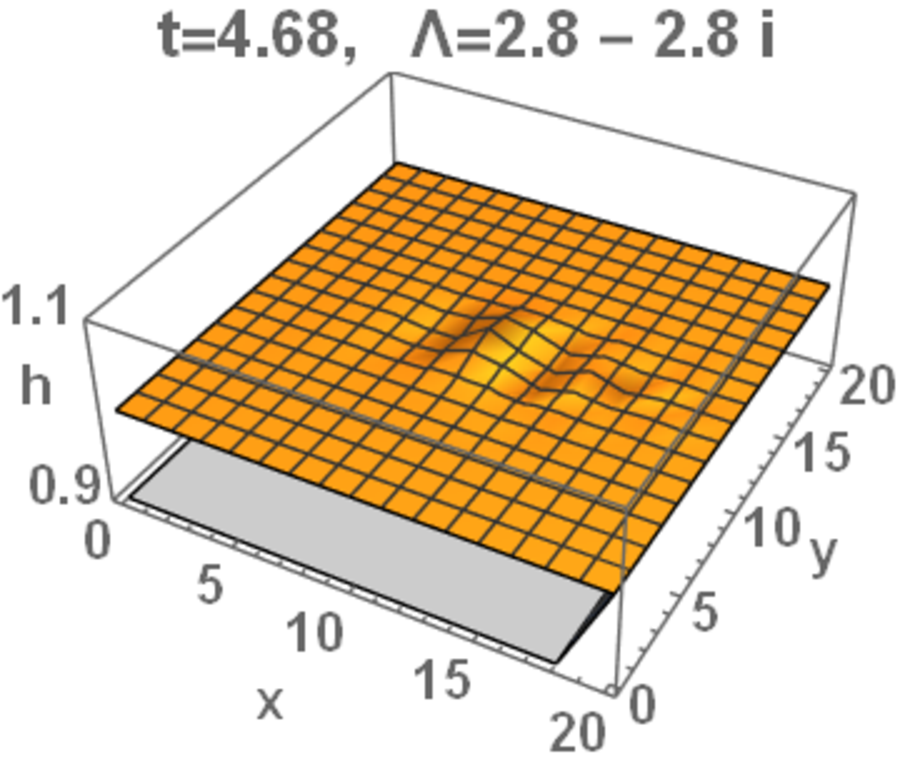}
\includegraphics[width=4cm]{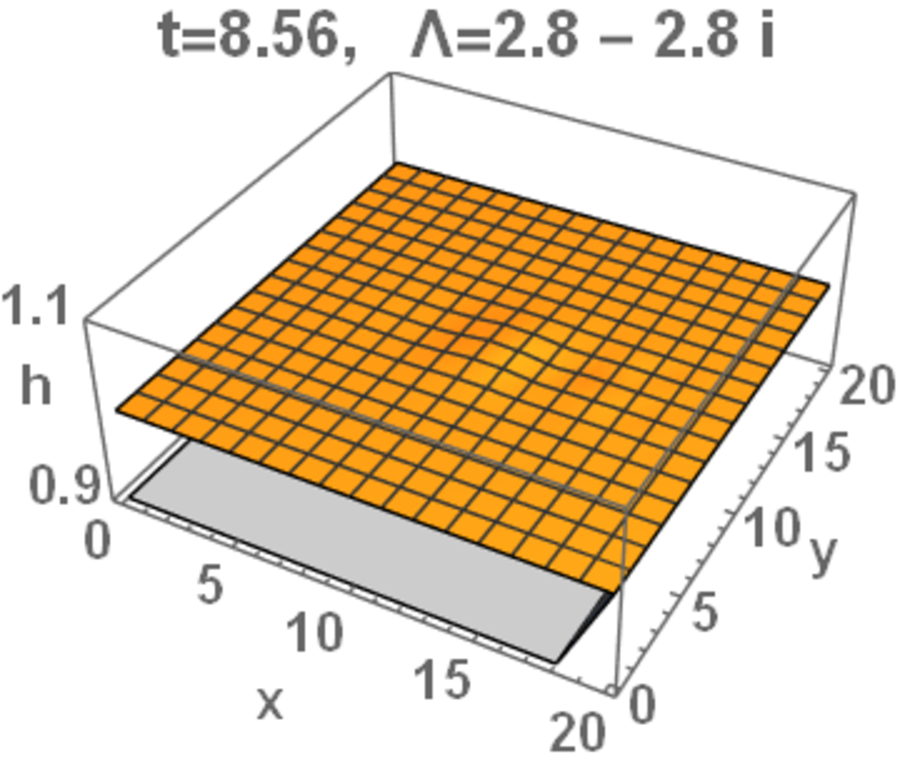}
}
\hspace{0.5cm}
\parbox[]{4.cm}{
\includegraphics[width=4cm]{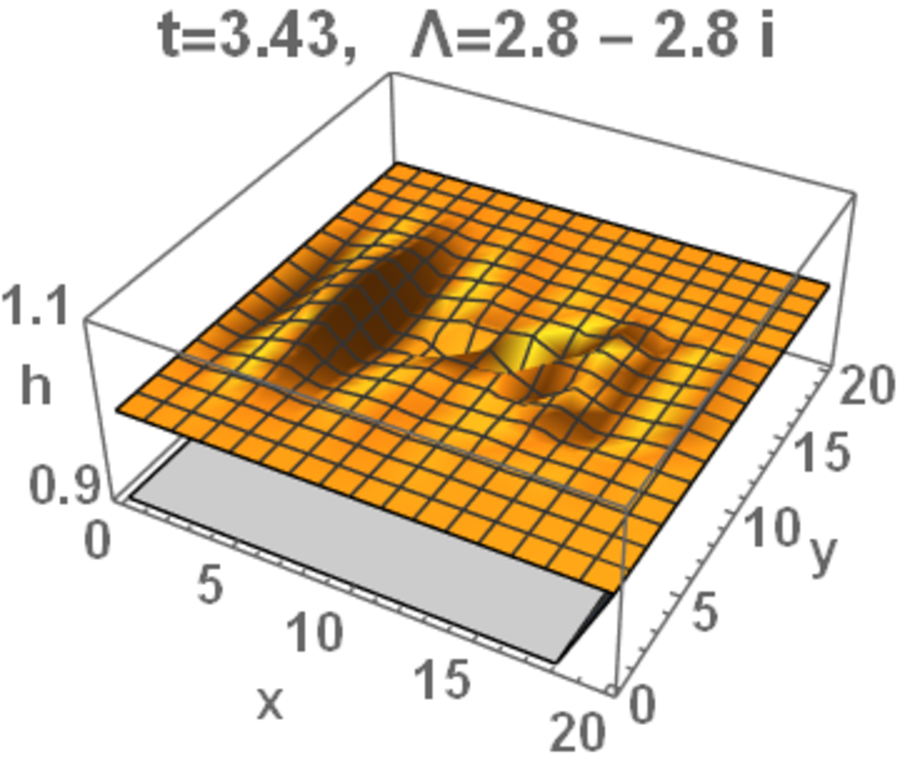}
\includegraphics[width=4cm]{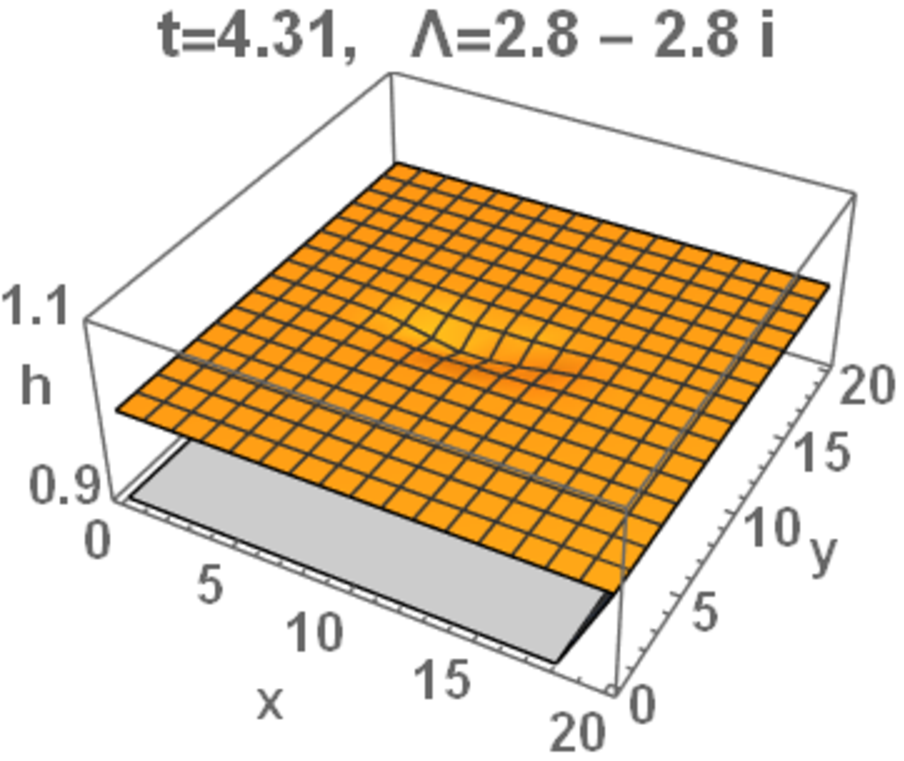}
\includegraphics[width=4cm]{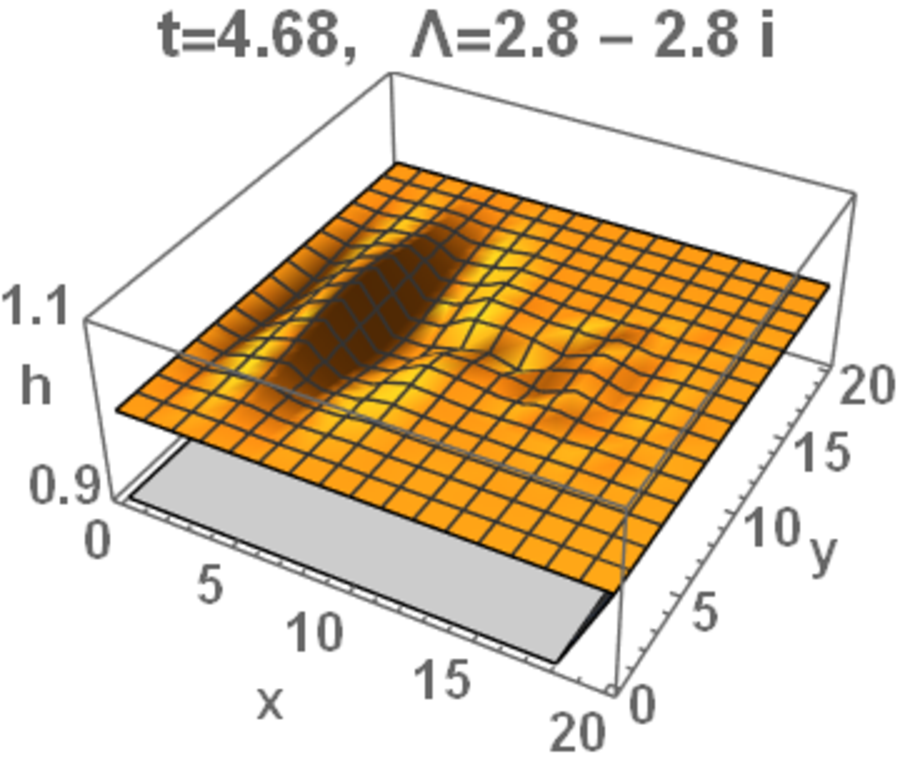}
\includegraphics[width=4cm]{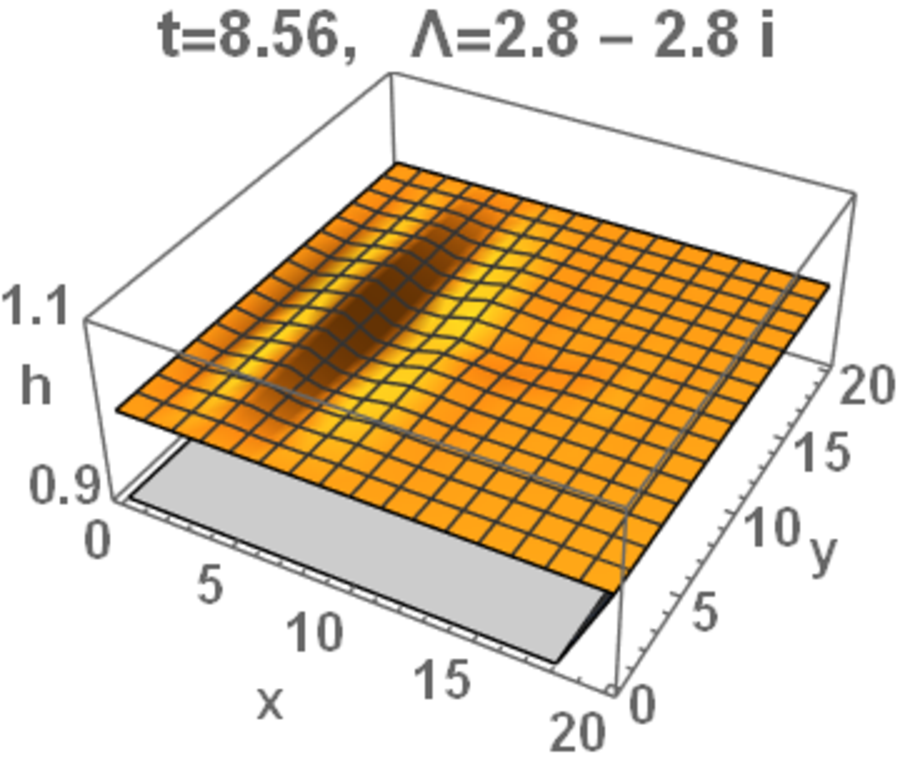}
}
}
\caption{\label{stripes3} Time steps of the two-dimensional set (\ref{sys}) at 1/10 of the initial elevation of figure~\ref{stripes1} for $G=1$, $\Gamma=0.5$, $H=2$ with external current $D_x=-6 \sin{\omega t}, D_y=0$ and $\omega=10$ without (left) and with (middle,right) tilted ground. The right column has been chosen with $D_x=-6.5$ being in the unstable regime for untilted bottom. The wavelengths of linear response are given above for comparison.}
\end{figure}

\end{widetext}

Finally we plot the ripple formation in x-direction in figure~\ref{stripes3} 
for external diffusion current $D_y=0$ and $D_x=-6$ just below the border to the unstable region. The appearance of ripples follows the external frequency. A linearly tilted bottom suppresses the ripple formation as seen in the middle column. In the right we enhance the external current and see that the ripples are more suppressed in the area where the bottom is approaching the surface.  

\subsection{Hexagonal structures}

To complete the discussion we are looking now for the hexagonal structure in some special cases. Neglecting the external beam we obtain from (\ref{sys1}) the solution 
\be
B_i=F_i,\, i=1,2,3.
\ee
Choosing specifically $F_2=F_3=0$ we have either
\be
&&B_1=F_1,\, B_2=B_3=0\nonumber\\
&&A_1=-{(G+\Gamma) F_1\over 2 H},\, A_2=A_3=0 
\ee
with the six growth rates from stability analysis (\ref{stab})
\be
\lambda=\bigl \{0,-2 H, -H\mp |(G+\Gamma) F_1\pm 4 H^2 | \bigr\}
\ee
or
\be
&&B_1=F_1,\, B_2=B_3=0\nonumber\\
&&A_1=-4 H,\, A_2=A_3 \mp\sqrt{-2 (G+\Gamma) F_1+16 H^2},\,{\rm or}\nonumber\\
&&A_1=4 H,\, A_2=A_3 \mp\sqrt{2 (G+\Gamma) F_1+16 H^2}
\ee
with the growths rates
\be
\lambda&=&\bigl \{ -H\mp\sqrt{\frac 3 2 (G+\Gamma) F_1+9 H^2},\nonumber\\
&&-H\mp\sqrt{(G+\Gamma) F_1+9 H^2},\nonumber\\
&&-H\mp\sqrt{-\frac 1 2 (G+\Gamma) F_1+9 H^2}\bigr\}.
\ee
None of these sets of growth rates can be simultaneously smaller zero. Therefore no stable hexagonal structure can appear without external beam.

In the other special case of no structure at the ground $F_1=F_2=F_3=0$ we can find the solution
\be
&&B_1=B_2=B_3=0\nonumber\\
&&A_3=-{1\over 2c_3}(A_1A_2-\sqrt{(A_2^2-4 c_1c_3)(A_1^2-4 c_2c_3}),\,{\rm or} \nonumber\\
&&A_1=\pm 4 H, A_2=\pm 4 H,A_3=\pm 4 H.
\ee
This means that no structure at the surface $B_i=0$ appears.
The second part of solutions lead to growth rates $\lambda=(-4 H,2 H,...)$ which shows unstable behaviour. 
Therefore only for special shaping of the ground we might expect hexagonal structure due to external beams. This could be analyzed further.

\section{Summary}

A model for time evolution of the liquidized metal is developed by hydrodynamic considerations. The Navier-Stokes equation together with the boundary of a variable ground are simplified by shallow-water approximations. We consider explicitly the effect of viscosity, surface tension and a fluid flow induced by an external laser or particle beam. To that end, two different models are presented which allow to describe the induced surface current on the surface. The resulting coupled equations for the height and two-dimensional velocity obey conservation laws of mass and momentum. It turns out that the gravitation and surface tension appear by a characteristic potential analogously to conservative forces. The viscosity modifies the momentum current density and leads to a damping term proportional to the spatial gradient of the velocity. Alternatively we could formulate it as a modification of the effective velocity with which the mean momentum is transported. The shape of the bottom contributes to the momentum balance only by two possibilities: either by a coupling to surface tension and gravitation or by coupling to the viscosity for spatial-dependent velocities. This underlines the nontrivial intrinsic interplay of surface tension, gravitation and viscosity. The presented model is discussed in the context of ripple formation in laser material processing, but is of rather general form and applies to other situations that involve the impact of a particle or laser beam on a liquid bath.

The linear stability analysis provides parameter ranges for viscosity and surface tension where stable or unstable oscillations can appear. The oscillating instability is shown to give rise to stripe structures. A minimal wavelength is identified where unstable oscillating behaviour can appear. The wavelength as function of frequency provides evanescent waves with wavelengths strongly dependent on the external current. For the ideal free case we have just gravitational waves. These waves becomes strongly modified by a combination of viscosity, surface tension, gravitation, and external current. A time periodic external beam creates further subregions of oscillatory instability which are determined by Floquet theory.

The weak nonlinear stability analysis shows that stripe or hexagonal structures can only appear if an external beam is present. The dependence of stripe orientations on the angle of incident beam to the surface is derived and a minimal incident angle is reported where stripe structures are possible. Due to surface roughness this incident angle is nonzero even for perpendicular impact. The stripe orientation angle is further restricted by the growth rates of the structure. The stability analysis provides a strong dependence of the stability of stripe structure on the amplitude compared to the external current. A maximal ratio of amplitude to current is reported only below which stripe structures can appear. We do not see a dependence of the stability on surface tension or gravitation but on the viscosity. Hexagonal structures are shown to be possible only if an external beam and a structured bottom is present. Since the shaping of the latter is beyond the considered experimental case this analysis is not followed further here.

\appendix

\section{Contribution of external current to surface distortion\label{curr}}

We develop two simplistic models suited for the distortion of the surface under the influence of an external beam or laser impact. We consider recoil models assuming that by the impact on the surface one side of the induced momentum is absorbed by the material and the other side gives rise to a surface current. In this sense we call it recoil models. Due to matter conservation we have to connect this resulting surface currents in x and y-direction
\be
J_x&=&-D_x \partial_x h,\quad
J_y=-D_y \partial_y h
\label{surfJ}
\ee
with the height in (\ref{sys}) as
\be
\dot h=...-\nabla\cdot \V J=...+D_x \partial_x^2 h+D_y\partial_y^2 h.
\ee
The angular dependence on the incident angle and the considered geometries will be absorbed in the diffusion coefficients $D_x,D_y$.

\subsection{Collisional recoil model}

\begin{figure}[h]
\includegraphics[width=6cm]{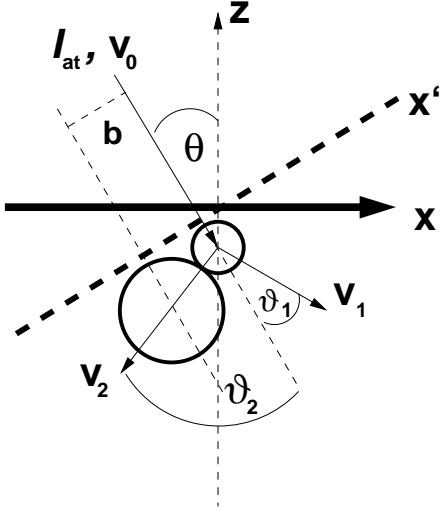}
\caption{\label{collision}The geometry of incoming sphere with velocity $v_0$ from a current $I_{\rm at}$
  colliding a sphere in the material under impact parameter $b$. Due to surface roughness and deformation of the surface by external beam the surface becomes tilted from $x$ to $x'$ direction blocking half of impact parameter. }
\end{figure}

We consider the collision of a project sphere $1$ with velocity $v_0$ to a target $2$ at
rest as illustrated in figure \ref{collision}.

 Let first look into the x-direction where the angle of recoil of the target is given by the impact parameter and
the sum of the two radii $ \sin\vartheta_2=b/R_{12}$. The elementary momentum and
energy conservation of this billiard model reads
\be
x p_2^2&=&p_0^2-p_1^2
\nonumber\\
\begin{pmatrix}p_0\cr 0\end{pmatrix}&=&
p_1\begin{pmatrix}\cos\vartheta_1\cr -\sin\vartheta_1\end{pmatrix}+
p_2\begin{pmatrix}\cos\vartheta_2\cr \sin\vartheta_2\end{pmatrix}
\ee
with the mass ratio $x=m_1/m_2$. Equating $p_1^2$ from the first and second line
yields the velocity of the target
atom 
\be
v_2={2 x\over 1+x} |\cos\vartheta_2| v_0
\ee
which gives the angular distribution of recoil velocities.  Each such ion creates energy-dependent atomic recoils $F_i(E)$ . Provided the atoms are present with relative concentration $c_i$ the total atomic recoil beam parallel to the surface reads 
\be
J_i= |\cos\vartheta_2| \sin{(\vartheta_2-\Theta)}I_{\rm at} \cos{\Theta} f_i c_i
\label{ji}
\ee
with $f_i={2 x_i\over 1+x_i}F_i(E)$ and we considered that the incoming beam
is $I_{at} \cos{\Theta}$.
Since $\vartheta_2$ is given by the ratio of the impact
parameter to the sum of radii, we average over all considered impact
parameters. Using as the range all impact parameter corresponding to the angle $-\pi/2<\vartheta_2<\pi/2$ would yield zero since all symmetric recoils sum up to zero. In the next step we will consider only gradients of the surface. This surface is assumed to be deformed indicated by the dashed line and $x`$ coordinate. Then the left side of figure~\ref{collision} which will couple to positive curvatures is creating a surface current while the right side is inside matter and is absorbed. Therefore we average only about the positive impact parameters which means about half of the available space $0<\vartheta_2<\pi/2$. We obtain the total beam parallel to the
surface due to the (first) recoil
\ba
&\bar J_i=I_{at} f_i c_i [a_c\cos{(\Theta)}-a_s\sin{(\Theta)}]\cos{\Theta}
\label{current}
\end{align}
with $a_c=1/3$, $a_s=2/3$. Of course, this first collision will lead to a further collisions and so on forming a whole cascade which would change the factors $a_s$ and $a_c$ slightly. 

Now we consider that
due to the surface inhomogeneity the angle $\Theta\to \Theta+\gamma_x$ fluctuates with the
gradient of the surface height $\gamma_x=\arctan(\partial h/\partial
x)\approx \partial h/\partial x$. The y-direction can be analogously considered as above for the x-direction but with $\Theta\to\gamma_y$ since the incident angle is zero in this direction and we expand in first orders of $\gamma_x,\gamma_y$ such that beyond the constant current the deviation is (\ref{surfJ})
\ba
D_x&=J_0 (a_c \sin{(2\Theta)}+a_s \cos{(2\Theta)}]
\nonumber\\
D_y&=J_0\,a_s
\label{comp}
\end{align}
with $J_0=I_{\rm at}f_i c_i$.
These atomic recoil currents act to smooth the surface. Please note that we
obtain another angular dependence than \cite{CV96}, where
$D(\Theta)=\cos{(2 \Theta)}$ and the y-direction is not considered. 

Please note that these surface currents couple to the second spatial derivative of the surface and are present only with corresponding roughness of the surface. As symmetry check we see that for perpendicular incident beams $\theta=0$ both surface directions couple equally $D_x=D_y=J_0 a_s$. For parallel impact we have the same value but different signs $D_x=-D_y=-J_0 a_s$.

\subsection{Surface impingement model}

In the last collision model we first describe the collision cascade as deterministic and then expand the final result in first-order gradients.
In contrast, the authors \cite{DAB07} have considered another model of expanding in first-order fluctuating gradients but relating the normal of the surface to a fixed geometrical incident angle.  Let us briefly outline the main idea of this model \cite{DAB07}.

\begin{figure}[h]
\includegraphics[width=8cm]{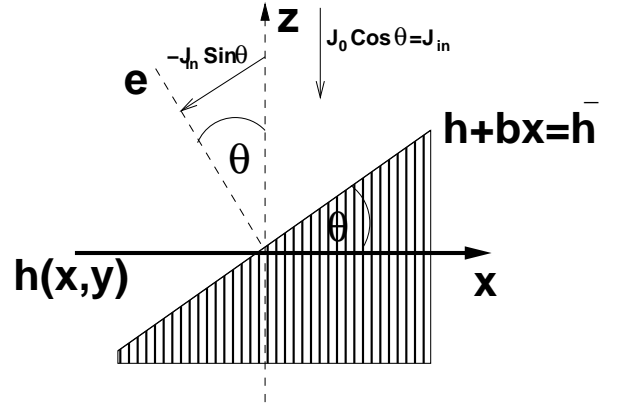}
\caption{\label{flow}The geometry of incoming impingement on the surface $J_0 \cos\theta $ split to the surface parallel contribution where $b=\tan \theta$.}
\end{figure}

The net surface impingement of an incoming current is $J_0\cos \theta$ with the incident angle $\theta$ of incoming beam to a plane parallel to $x-z$ plane as illustrated in figure~\ref{flow}. It is assumed that the surface will be tilted according to the incoming beam such that $\bar h=h+b x$ where we abbreviate $b=\tan\theta$. The net surface current is then
\be
J=-J_0 \sin\theta \cos\theta=-{J_0\over 2}\sin 2\theta.
\label{i1}
\ee
This surface current in the incoming plane is distributed in $x$ and $y$ direction according to the angle determined by the surface roughness
\be
\tan \varphi&=&{\p{y}\bar h\over \p{x} \bar h}\nonumber\\
\cos\varphi&=&{\p{x} \bar h\over |\nabla_2 \bar h|}\nonumber\\
\sin\varphi&=&{\p{y} \bar h\over |\nabla_2 \bar h|}.
\label{i2}
\ee
Geometrically we can express 
\be
\cos\theta&=&\V e\cdot \V e_z={1\over \sqrt{1+(\nabla_2\bar h)^2}}\nonumber\\
\ee
With $\nabla_2 \bar h=\p{x} h+b+\p{y}h$ we expand in first-order derivatives of the surface to get
\be
\cos\theta&=&{1+b^2-2 b \p{x}h\over (1+b^2)^{3/2}}\nonumber\\
\sin\theta&=&{b+b^3+\p{x}h\over (1+b^2)^{3/2}}\nonumber\\
\cos\varphi&=&1\nonumber\\
\sin\varphi&=&{\p{y} h\over b}.
\ee
This leads with (\ref{i1}) and (\ref{i2}) to
\be
\p{x}J_x&=&\p{x} J \cos\varphi=-J_0{(1-b^2)\over (1+b^2)^2}\p{x}^2 h 
\nonumber\\
\p{y}J_y&=&\p{x} J \sin\varphi=-J_0{1\over 1+b^2}\p{y}^2 h. 
\ee
The factors in (\ref{surfJ}) we obtain by rewriting $b=\tan\theta$ 
\be
D_x&=&{J_0\over 2} \cos 2\theta (1+\cos 2\theta)
\nonumber\\
D_y&=&{J_0\over 2} (1+\cos 2\theta).
\label{comp1}
\ee
Again we comment that this surface current couples on the derivatives of the surface and therefore the roughness of the surface. For perpendicular beam direction $\theta=0$ we obtain symmetric coupling $D_x=D_y=J_0$ and for parallel beam there is no surface current since this model assumes a tilting of the surface due to the beam which is not happening for parallel beams.

\section{Momentum balance\label{mbalance}}

Using (\ref{vel2}) in (\ref{vel2a}) one obtains
\ba
\p{t}[(h-f)\V u]&=-\V u\{ \nabla_2\cdot[(h-f)\V u]\}-(h-f)(\V u\cdot\nabla_2)\V u\nonumber\\
&
-(h-f)\nabla_2( G h \!+\! \Gamma  \nabla_2^2 h \!+\! 2H \nabla_2\cdot \V u).
\end{align}
The first line can be written into two forms
\be
&\V u \p{j}[(h-f)u_j]-(h-f)u_j\p{j} \V u=-\p{j}[(h-f)u_j\V u]
\nonumber\\
&=-u_j\p{j}[(h-f)\V u]-\V u(h-f)\p{j}u_j
\ee
where the first form is just part of the momentum current density (\ref{stress})
and the second form contributes to the substantial derivative (\ref{pbal1})
as well as part of the source term (\ref{s1}).

The gravitational $G$ and surface tension $\Gamma$ part of the second line can be written as the negative gradient of the potential (\ref{V}) as seen by inspection.

The remaining viscosity part $~2H$ can be written
\be
-(h-f)\p{i}\p{j} u_j
=-\p{j}[(h-f) \p{i}u_j]+\p{j}(h-f)\p{i} u_j
\ee
where the first part gives the contribution to the momentum current density (\ref{stress}). Using $\p{i} u_j=\p{j} u_i$ due to the curl-free condition (\ref{eq:u0v0}) we can rewrite the second part as
\ba
&\p{j}(h\!-\!f)\p{i} u_j=\p{j}(h\!-\!f)\p{j} u_i
\nonumber\\
&=\p{j}\ln (h\!-\!f) \p{j}[(h\!-\!f)u_i]-u_i \p{j}\ln(h\!-\!f)\p{j}(h\!-\!f).
\end{align}
The first part renormalizes the substantial derivative in (\ref{pbal1}) and the second part contributes to the source term (\ref{s1}).



\end{document}